\documentclass[12pt]{article}
\usepackage{graphics}
\usepackage{graphicx}
\DeclareGraphicsExtensions{.pdf}
\usepackage{float} 
\usepackage{amsmath}
\usepackage{slashed}
\usepackage{amsfonts}   
\usepackage{amssymb}

\usepackage{color}
\begin{document}
\date{}
\title{{\bf{\Large Fragmentation and defragmentation of strings in type IIA and their holographic duals}}}
\author{
 {\bf {\normalsize Dibakar Roychowdhury}$
$\thanks{E-mail:  dibakarphys@gmail.com, dibakar.roychowdhury@ph.iitr.ac.in}}\\
 {\normalsize  Department of Physics, Indian Institute of Technology Roorkee,}\\
  {\normalsize Roorkee 247667, Uttarakhand, India}
\\[0.3cm]
}

\maketitle
\begin{abstract}
We probe type IIA geometries with \emph{folded} semiclassical strings those encode the strong coupling dynamics associated with long operators in a class of SCFTs pertaining to spacetimes whose dimension vary from $6d$ down to $2d$. We particularly focus on folded string configurations those are extended through stack of flavor $ Dp $ ($ p=6,8 $) branes localized along the internal manifold. Considering some specific examples within the realm of $ \mathcal{N}=2 $ linear quivers, we show that the associated string spectrum exhibits a pole as the string approaches flavor D6 branes. We identify this as a \emph{fragmentation} of long operators in the dual $ \mathcal{N}=2 $ SCFTs. On the other hand, a similar analysis for $ \mathcal{N}=(1,0) $ SCFTs in $6d$ as well as $ \mathcal{N}=(0,4) $ SCFTs in $ 4d $ reveals a set of new dispersion relations at strong coupling. Based on semiclassical calculations, we set an argument that explains either of these cases.
\end{abstract}
\section{Overview and Motivation}
\subsection{Type IIA dualities}
The duality between type IIB strings on $ AdS_5 \times S^5 $ and $ 4d $ $ \mathcal{N}=4 $ SCFTs \cite{Maldacena:1997re} has been tested on various grounds.  One of the nontrivial checks of the above correspondence comes from studying the spectra associated with long (\emph{folded}) string configurations on $ AdS_5 \times S^5 $ \cite{Gubser:2002tv} where the anomalous dimension associated with gauge theory operators with large spin $ S (\gg \sqrt{\lambda})$ have been identified with excited spinning string states on $ AdS_5 \times S^5 $. These single spin solutions were further generalized for multispin states \cite{Frolov:2003qc}-\cite{Frolov:2003xy} and a precise map between these string states and single trace operators in gauge theory was established \cite{Beisert:2003xu}. 

Following the same spirit as above, one can in fact ask whether similar checks can be performed for a large class of type IIA dualities those were put forward (as well as tested on various grounds) during past one decade until very recently. This is altogether a nontrivial question whose complete answer is not known yet. However, some part of it can be addressed. The purpose of the present paper is therefore to build up a systematic algorithm to pursue this quest by taking some specific examples of quantum field theories whose strong coupling behavior can be explored by studying a class of type IIA supergravity solutions in $ 10d $. In particular, these are the quantum field theories (with different amount of SUSY) those can be classified as supersymmetric conformal field theories (SCFTs) living in diverse dimensions starting from $ 6d $ down to $ 2d $.

Below we briefly outline each of these theories, their dual (type IIA) supergravity counterpart and the associated Hanany-Witten brane set up.

The first example we consider is that of $ \mathcal{N}=2 $ linear quivers in $ 4d $ \cite{Gaiotto:2009we}. These quivers have their dual supergravity description in $ 10d $ known as
Gaiotto-Maldacena (GM) backgrounds \cite{Gaiotto:2009gz}. These spacetimes can be built up systematically, by introducing potential functions \cite{Sfetsos:2010uq}-\cite{Maldacena:2000mw} (associated with a unique charge density) and thereby considering a linear superposition of these functions those are dual to $ \mathcal{N}=2 $ quivers with richer structure \cite{ReidEdwards:2010qs}-\cite{Nunez:2019gbg}. The corresponding field contents of $ \mathcal{N}=2 $ quivers are realised as a low energy limit of NS5-D4-D6 brane set up. The $ N_c $ D4 branes stretched between two NS5 branes (along the holographic/field theory direction ($ d=\eta) $) corresponds to adding a $ SU(N_c) $ color group with $ \mathcal{N}=2 $ quivers. On the other hand, the presence of $ N_f $ D6 branes between two consecutive intervals of NS5 branes corresponds to adding a $ SU(N_f) $ flavour group such that $ N_f=2N_c $ is satisfied for each node of the quiver \cite{Lozano:2016kum}.

The second example we consider is that of $ 6d $ $ \mathcal{N}=(1,0) $ SCFTs dual to (massive) type IIA supergravity solutions with an $ AdS_7 $ factor \cite{Apruzzi:2013yva}-\cite{Bergman:2020bvi}. These $ \mathcal{N}=(1,0) $ quivers are realized as a low energy limit of Hanany-Witten set up \cite{Hanany:1996ie}-\cite{Brunner:1997gk} which corresponds to NS5-D6-D8 brane intersections in $ 10d $. Within this brane set up, the number ($ N_c $) of D6 branes stretched between two NS5 branes counts the color content of the quiver associated to a particular node. On the other hand, the number ($ N_f $) of D8 branes between two consecutive intervals (of NS5 branes) counts the number of flavour degrees of freedom associated with a particular node of the $ \mathcal{N}=(1,0) $ quiver.  Like in the previous example of $ \mathcal{N}=2 $ quivers, one has to satisfy the constraint, $ N_f =2 N_c $ for each node of the quiver.

The third and the final example we consider is that of $ \mathcal{N}=(0,4) $ linear quivers in $ 2d $ and their type IIA duals with an $ AdS_3 $ factor. These quivers were constructed very recently in a series of papers \cite{Lozano:2019emq}-\cite{Filippas:2020qku}. $ \mathcal{N}=(0,4) $ SCFTs can be realized as IR fixed point of some RG flow starting from a UV complete QFT with $ \mathcal{N}=(0,4) $ SUSY. In a Hanany-Witten set up, these quivers can be realized as an intersection of NS-D2-D4-D6-D8 branes in $ 10d $. The D2 and D6 branes source two long array of color nodes in the $ \mathcal{N}=(0,4) $ quiver. On the other hand, the D4 and D8 branes stand for the flavour nodes of the quiver. The color nodes in the quiver are connected to each other through $ (4,4) $ and $ (0,4) $ hypermultiplets as well as $ (0,2) $ Fermi multiplets. On the other hand, the color groups are connected with flavour groups through Fermi multuplets. 

Each of these SCFTs shows a unique feature. Corresponding to each class, these superconformal quivers can be categorized following different combinations of color and flavour groups.  In our analysis, we will be mostly concerned with two types of finite quivers namely the ``single kink" (Fig.\ref{uluru}a) and quivers with a \emph{plateau} region (Fig.\ref{uluru}b). Being finite, each of these quivers can be associated with long operators with finite energy spectrum.
\subsection{Summary of results}
\begin{figure}
\includegraphics[scale=.79]{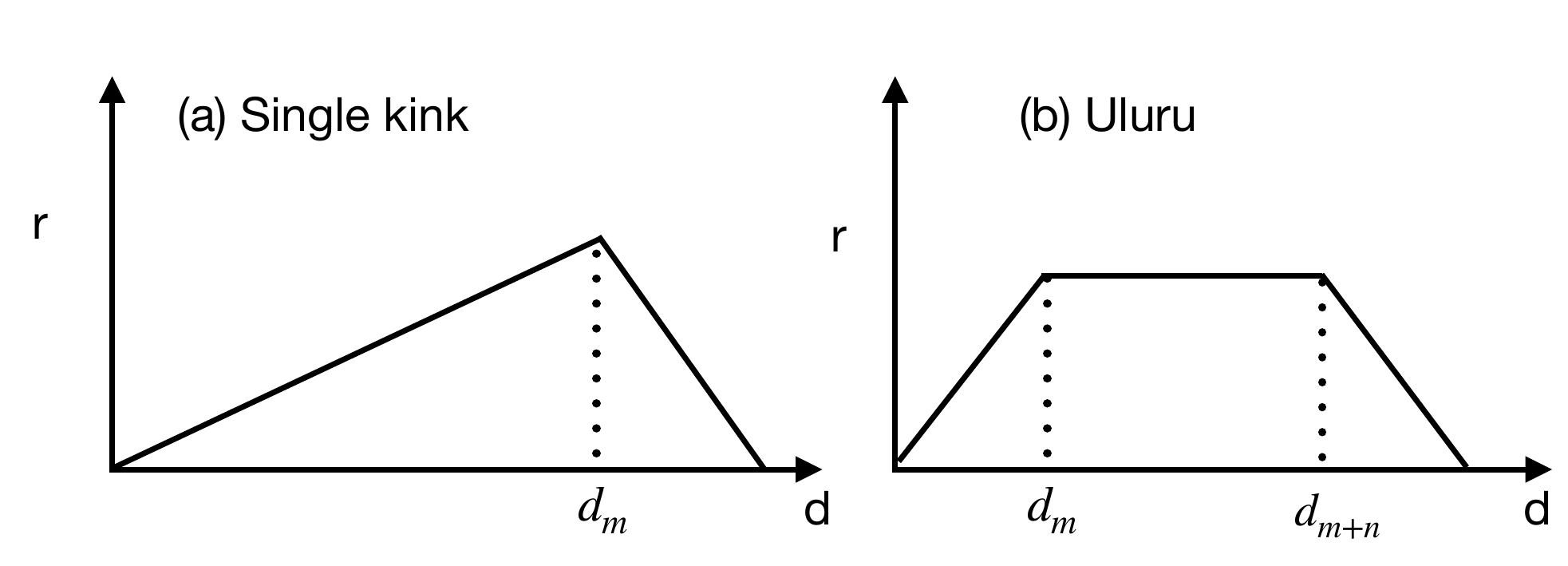}
  \caption{We plot the rank ($ r $) of the gauge group against the holographic direction ($ d $) which in the dual string theory description is identified as one of the coordinates associated with the internal manifold. (a) Single kink profile corresponds to a quiver with linearly increasing rank that is closed by adding a $ SU((m+1)N_f) $ flavour group at the end of the long chain. In the above figure this correspods to placing flavour branes at $ d=d_m $ where $ m $ is a positive integer that increases by one, (b) Uluru profile corresponds to associating flavour branes at two distinct points ($ d_m $ and $ d_{m+n} $) along the nodes of the quiver and thereby changing the slope at $ d_m $ and $ d_{m+n} $.} \label{uluru}
\end{figure}

To understand the key difference between long operator states for each of these SCFTs we refer to Fig.\ref{uluru}. Long operators in each of these gauge theories are dual to semiclassical \emph{folded} strings extended through flavour Dp branes located along the (holographic) axis ($ =d $) of the internal space of type IIA solution. As the string passes through different flavour branes, it experiences a potential that affects the energy of the string which in turn is also reflected in the corresponding spectrum of heavy operators in the dual theory.

\begin{figure}
\includegraphics[scale=0.82]{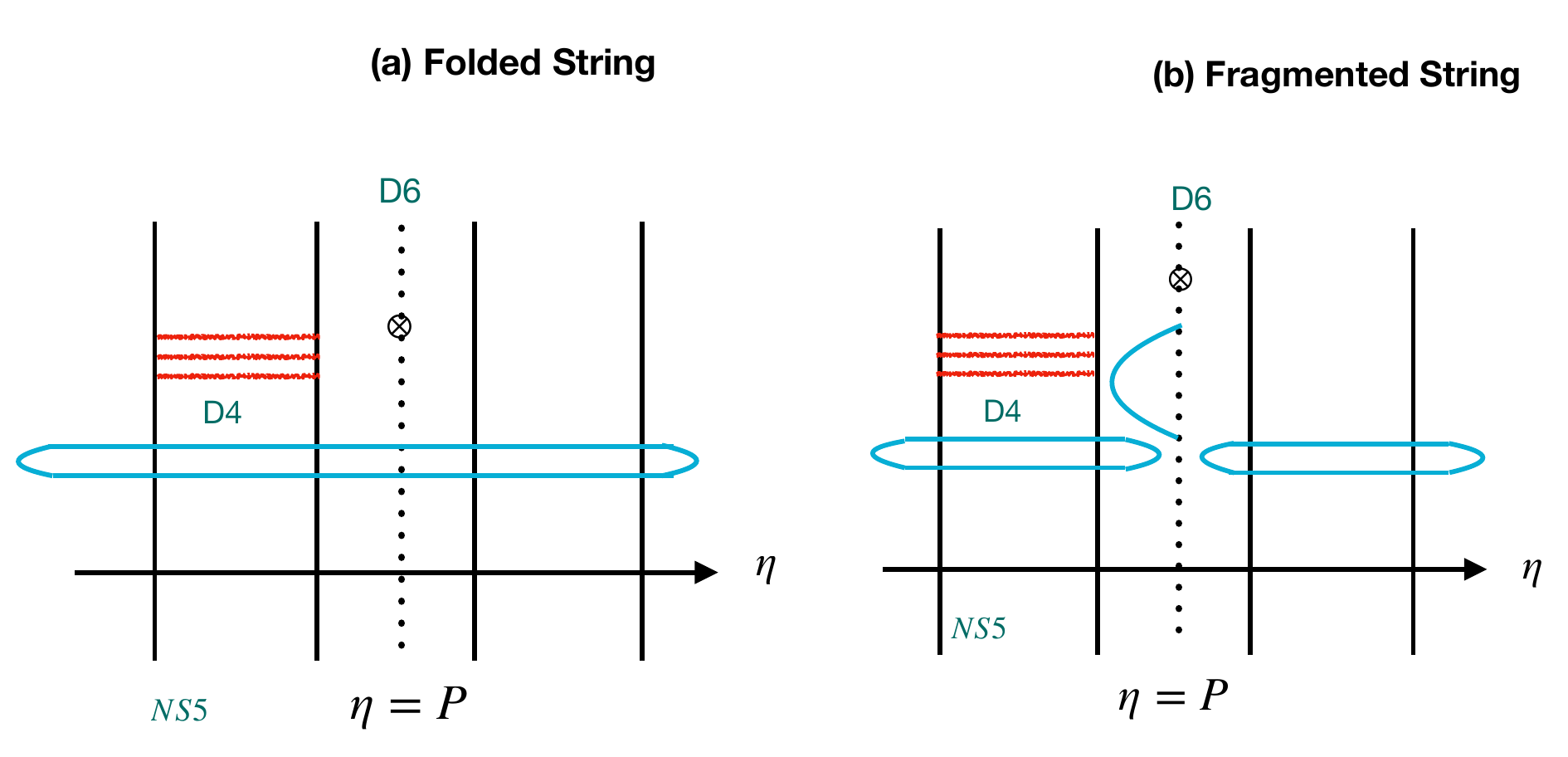}
  \caption{(a) Long folded string passing through flavour D6 branes in a Hanany-Witten setup corresponding to $ \mathcal{N}=2 $ linear quivers, (b) Fragmented strings dual to $ \mathcal{N}=2 $ SCFTs in $ 4d $.} \label{folded-string}
\end{figure}

Below we elaborate on this in detail.\\
$ \bullet $ $ 4d $ $ \mathcal{N} =2$ linear quivers turn out to be very special. Long folded strings in the dual GM geometry ``feel" an infinite potential barrier as it approaches flavour D6 branes near the singularity at $ \eta \sim P \sim d_m $ (Fig.\ref{folded-string}).  This causes an infinite shift in the corresponding energy spectrum of these strings. Using a reverse engineering method this can be explained as follows: the folded strings can actually never pass through flavour D6 branes - as a result if one starts with long folded string configurations (those cover the entire GM geometry) they are naturally \emph{fragmented} into smaller string segments those are floating around flavour D6 branes located at $ \eta \sim P \sim d_m$ (Fig.\ref{folded-string}b). The energy associated with these smaller string segments can be obtained following a proper regularization scheme. 

Below, we summarise the energy associated with smaller string segments corresponding to each of these quivers\footnote{ For the details of the derivation the reader is refereed to Section \ref{strings through flavor branes}.} as depicted in Fig.\ref{uluru}
\begin{eqnarray}
E_S \sim \begin{cases}
      J^{1/4}&  (\text{Fig.\ref{uluru}a})\\
      J^{1/3} & (\text{Fig.\ref{uluru}b})
    \end{cases} 
\end{eqnarray}
where $ J $ is the generator of rotation associated to $ S^2 $ of the internal manifold.

One can perform similar calculations in the absence of rotation and consider simply extended strings those are passing through flavour D6 branes. Once regularised properly, these calculations yield a finite energy\footnote{See Section \ref{revisiting strings in single kink and uluru} for the details of the derivation.} corresponding to those smaller string segments floating around flavour D6 branes around $ \eta \sim P\sim d_{m} $.\\
$ \bullet $ A similar calculation for $ 6d $ $ \mathcal{N}=(1,0) $ SCFTs and $ 2d $ $ \mathcal{N}=(0,4) $ SCFTs, on the other hand, yield a finite answer. We interpret this as an artefact of the presence of a \emph{regular} potential profile near the flavour D8 branes at $ d \sim d_m $. Therefore long strings can smoothly pass through these flavour D8 branes without showing any divergences in the associated spectrum. As in the case of $ \mathcal{N}=2 $ quivers, these results can be generalised in the presence of non-zero charge $ J $. 

Below we summarise the spectrum for each of these quivers:
\begin{eqnarray}
E_S -J = \begin{cases}
    g  Q^{\frac{1}{p}}_{D6}&  $$ \mathcal{N}=(1,0) $$ \\
    f  Q^{\frac{1}{p+1}}_{D6}& $$ \mathcal{N}=(0,4) $$ 
    \end{cases} 
\end{eqnarray}
where we identify $ Q_{D6} $ as the Page charge associated with color $ D6 $ branes. Here, $ g $ and $ f $ are the constants of proportionality whose values are unique to the choice of the quiver\footnote{The constant of proportionality $ g $ has been introduced in (\ref{eee178}) in the context of $ \mathcal{N}=(1,0)  $ SCFTs in $ 6d $. Our analysis reveals that the value of $ g $ depends on the choice of the particular quiver. For example, quivers without a plateau (\ref{e146}) correspond to $ g = \frac{7}{\sqrt{2N}}$ (see (\ref{eee174})) where $ N $ is the number of color nodes associated to the quiver. On the other hand, for quivers with a plateau (\ref{e148}) we find $ g= \frac{347}{576n}$ (see (\ref{eee177})) where $ n $ is the rank of the associated flavour group of the quiver. In a similar note, we define the proportionality constant $ f $ for $  \mathcal{N}=(0,4) $ SCFTs in $ 2d $ (see (\ref{eee224})). The value of $ f $ for different choices of the quivers has been summarised in (\ref{e219}) where $ \nu $ stands for the rank of the corresponding color group.}.

Notice that, here $ p $ is an integer which turns out to be quite unique corresponding to the choices of the quivers in each these SCFTs. It turns out that, $ p=2 $ corresponding to the single kink profile as depicted in Fig.\ref{uluru}a. On the other hand, it turns out be unity for those quivers which exhibit a plateau region as in Fig.\ref{uluru}b.

Finally, we conclude in Section 5 where we briefly outline the interpretation of the above findings in terms of the degrees of freedom associated with the dual SCFTs at strong coupling.
\section{Preliminaries}
The present Section, is intended to offer a general introduction to the basics of Hanany-Witten (HW) brane set up \cite{Hanany:1996ie}-\cite{Brunner:1997gk} that realizes the SCFTs (with different amount of SUSY) in diverse dimensions. We also discuss the supergravity limit of these brane intersections which forms the basis for the rest of the analysis of the paper. 

Although during the process of illustration, as a model, we consider $ \mathcal{N}=1 $ SCFTs however, a similar description equally applies to other two theories as well. These are the $ \mathcal{N}=2 $ SCFTs in $ 4d $ and $ \mathcal{N}=(0,4) $ SCFTs in $ 2d $ with respective modifications in the corresponding HW brane set up. For example, $ \mathcal{N}=2 $ SCFTs in $ 4d $ are realized as an NS5-D4-D6 brane intersections in $ 10d $ where D4s stand as color branes and D6s are the flavor branes. On the other hand, the HW set up for $ \mathcal{N}=(0,4) $ SCFTs in $ 2d $ has been discussed briefly below (\ref{eq201}).

The generic HW brane set comprises of an intersection of NS5-DN$ _c $-DN$ _f $ branes in $ 10d $ where $ N_c $ corresponds to the rank of the color group and $ N_f $ stands for the rank of the associated flavour group. We illustrate this with an example of $ \mathcal{N}=(1,0) $ SCFTs in $ 6d $.

Consider a brane configuration in which NS5 branes are separated which therefore gives rise to VEV ($ \langle \Phi \rangle $) to the scalars in the \emph{tensor} multiplet (which consists of a self dual two form ($ B_2 $) and a real scalar ($ \Phi $)) of $ \mathcal{N}=(1,0) $ theory. In this description, the so called tensionless strings stretched between two NS5 branes are actually ``massive" and hence decouple from the rest of the spectrum. 

Notice that, in this description, $ \langle \Phi_{i} \rangle $ provides the position of the $ i $th NS5 brane along the holographic axis ($ X^6 $). Clearly, the entity $ \langle \Phi_{i +1} \rangle -\langle \Phi_{i} \rangle $ measures a relative separation between the NS5 branes. Under such circumstances, the $ 6d $ world-volume theory is said to be described by an \emph{effective} linear quiver, where by the word effective we mean that the theory is described away from its UV fixed point in a RG flow.

\begin{figure}
\includegraphics[scale=0.73]{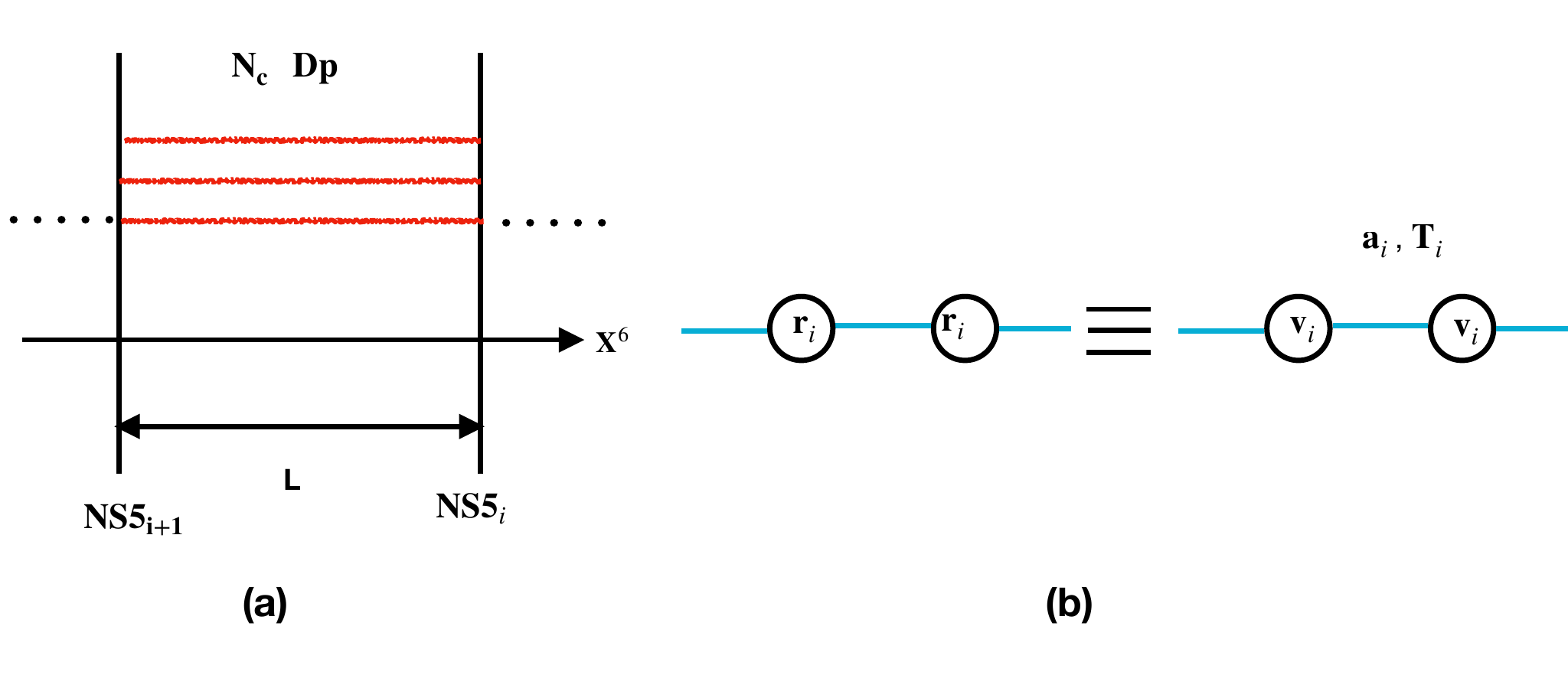}
  \caption{(a) Two NS5 branes (located at $ i $ and $ i+1 $ along the $ X^6 $ direction) are separated due to N$ _c $ color Dp branes placed in between, (b) Different vector multiplets ($ V_i $) connected through a link which is built out of tensor multiplets ($ T_i $).} \label{ns5}
\end{figure}

Consider a situation in which two NS5 branes (at positions $ i $ and $ i+1 $ (See Fig.\ref{ns5}a)) are separated by a distance $ L (\sim \langle \Phi_{i +1} \rangle -\langle \Phi_{i} \rangle)$ and $ N_c $ color D$ p $ ($ p=6 $) are stretched between them. As a result, the effective coupling on the world-volume of these D$ p $ branes is given by, $ g^2_{YM}=\frac{g_s l_s^{p-3}}{\langle \Phi_{i +1} \rangle -\langle \Phi_{i} \rangle} $. In other words, the VEV of scalars ($ \langle \Phi \rangle $) in $ p=6 $ dimensions plays the role of the effective gauge coupling for the D$ p $ brane world-volume theory.

Clearly, in the limit ($ L \rightarrow 0 $) when these NS5 branes are put on top of each other, the corresponding world-volume theory for color D6 branes becomes strongly coupled and the corresponding theory flows to a UV conformal fixed point that preserves $ \mathcal{N}=(1,0) $ SUSY in $ 6d $. This is precisely what we mean by the ``holographic" or the supergravity limit. The reason this is true is because the only algebra with $ 8 $ supercharges in $ 6d $ is the $ \mathcal{N}=(1,0) $ superconformal algebra. In this strongly coupled limit, the massless modes of the $ 6d $ theory are sourced due to ``tensionless" strings stretched between NS5 branes. 
\subsection{Linear quivers}
Let us now illustrate more on the definition of a quiver and its pictorial representations as in Fig.\ref{uluru}. A quiver essentially could be thought of as a collection of ``nodes" and ``arrows" which forms a useful way of describing the matter content of a supersymmetric conformal field theory. Let us illustrate on the above taking example of NS5-D6-D8 brane intersections in $ 10 d$ where $ N $ NS5 branes are separated by putting color D6s in between. 

The massless sector of the corresponding $ \mathcal{N}=(1,0) $ SCFTs consists of -\\
$\bullet $ Vector multiplets: $V_i \equiv (A_{\mu}  ,  \lambda_{\alpha}  , D)_{i}~ (i=1,\cdots , N-1)$ in the adjoint representation of $ U(r_i) $ where, $ r_i $ is the rank of the color (gauge) group for the $ i $th node.\\
$\bullet $ Hyper-multiplets: $ (h ,  \psi_{\dot{\alpha}} )_{i}~ (i=1,\cdots , N-2)$ in the bi-fundamental of the gauge group.\\
$\bullet $  Hyper-multiplets: $ (\tilde{h}^{a^i} ,  \tilde{\psi}^{a^i}_{\dot{\alpha}} )~(a^i=1,\cdots , f_i ~, ~\forall i=1, \cdots , N-1)$ in the fundamental of the gauge group where each $ h $ and $ \tilde{h} $ represents four real scalars and $ \psi, \tilde{\psi} $ represents left handed Weyl spinors and $ f_i $s are the flavour indices.\\
$\bullet $ Tensor multiplets: $T_i \equiv (\Phi , \chi_{\alpha} , B_{\mu \nu})_{i}\in $ NS5 $ (i = 1, \cdots , N) $ in the singlet representation of the gauge group where $ \chi_{\alpha} $ is the left handed Majorana spinor, $ \Phi $ is the real scalar and $ B_{\mu \nu} $ is the self dual two form as mentioned previously.\\
$\bullet $ Linear multiplets: $ (\Pi , C , \xi_{\dot{\alpha}})_i (i=1, \cdots ,N) $ where $ \Pi $ and $ C $ stands for the scalar multiplets and $\xi_{\dot{\alpha}}  $ are the left handed Weyl spinors.
\subsection{Constructing the quiver}
We now provide the basic nomenclature for constructing a quiver. As for example, consider $\mathcal{N}=(1,0)  $ SCFTs in $ 6d $. The world-volume theory on color D6 branes has a gauge group $ U(r_i)= U(1)\times SU(r_i) $ where the remaining $ U(1) $ mode eventually becomes massive and decouples from the rest of the spectrum.

Below, we enumerate essential steps in order to represent a quiver.\\
$ \bullet $ For the color (gauge) group $ SU(r_i)\in$ D6, pictorially we represent it by adding a round node (Fig.\ref{ns5}b) which characterizes a vector multiplet ($ V_i \in \mathcal{N}=(1,0)$) at the $ i $th site. Two such nodes are connected by a ``link" which is constructed out of the bi-fundamental matter $ (h ,  \psi_{\dot{\alpha}} )_{i}~(= a_i)$ together with the matter contents from tensor multiplets ($ T_i $).\\
$\bullet$ The presence of the flavor group $ SU(f_i) $ (where $ f_i $ counts the number of D8s) between two consecutive NS5 branes (located at $ i $ and $ i+1 $) one adds a box.\\
$\bullet$  The gauge anomaly cancellation requires to set, $ 2r_i -r_{i+1} -r_{i-1}=f_i$.

Let us now illustrate on these taking simple examples of ``single kink" quivers as well as quivers with a \emph{plateau} region. In this paper, we discuss these two quivers only.

\begin{figure}
\includegraphics[scale=0.8]{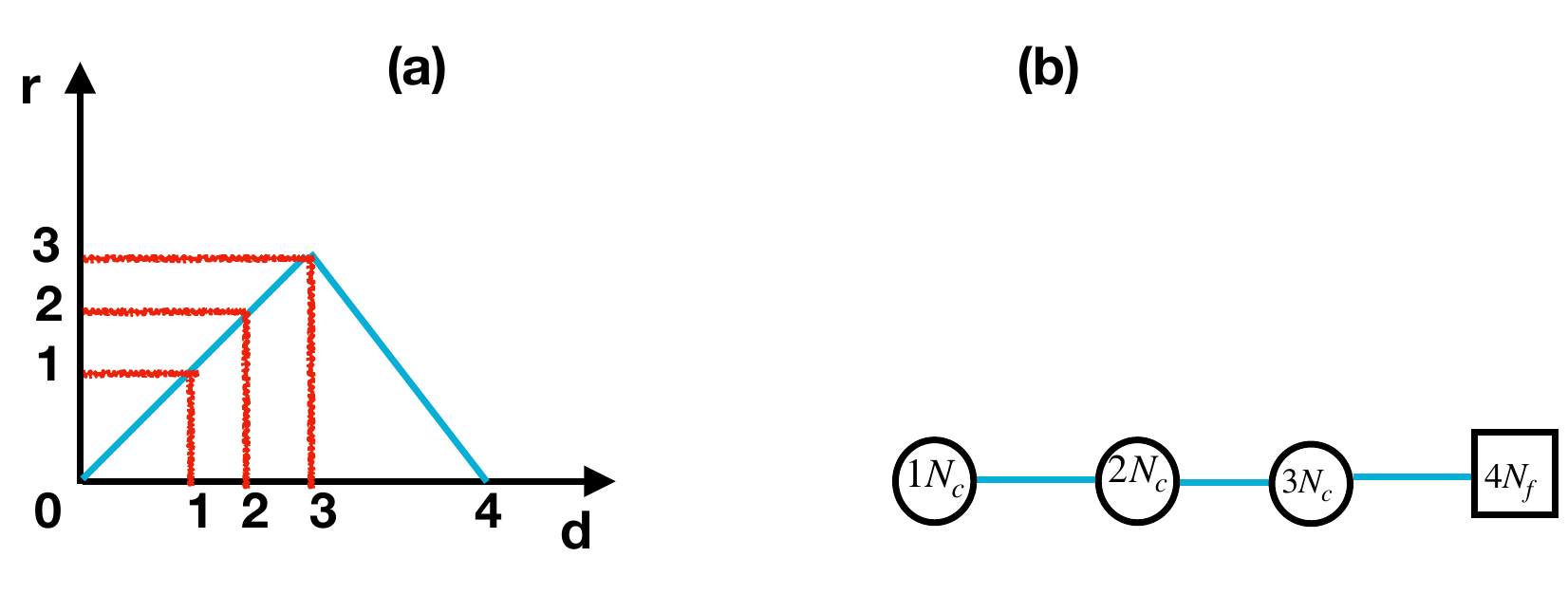}
\includegraphics[scale=0.8]{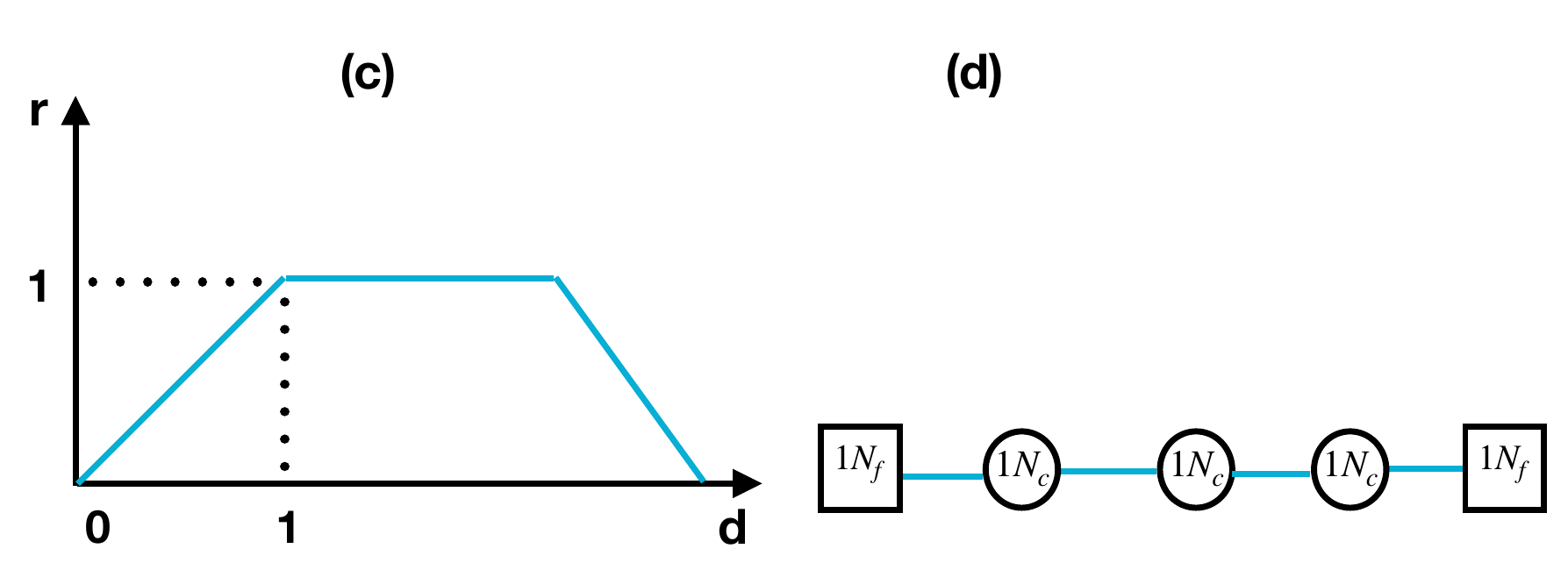}
  \caption{(a)-(b) Quivers with single kink, (c)-(d) Quivers with a plateau} \label{rank}
\end{figure}

In Fig.\ref{rank}b, we show a quiver with linearly increasing rank starting from $ 1N_c $ upto $ 3N_c $. The quiver is closed by placing flavor branes with $ SU(4N_f) $ flavor group. Notice that, at each node of the quiver, one satisfies the constraint that needs to be satisfied in order to cancel gauge anomalies. In the adjacent Fig.\ref{rank}a, we show a diagrammatic representation of this quiver in the ``holographic" limit where we plot the rank function ($ r $ in units of $ N_c $) against the holographic axis ($ d $). 

In Fig.\ref{rank}d, we show an example of a quiver with a plateau region. This quiver is closed by placing flavor nodes at the beginning and at the end of the color nodes. Clearly, the rank of the gauge group does not change between two flavor nodes which thereby gives rise to the plateau region (Fig.\ref{rank}c). In both the examples, we notice that the addition of flavor degrees of freedom changes the slope of the rank function.

\section{ 4d $\mathcal{N}=2 $ SCFTs and type IIA dual}
We start by considering (semi)classical stringy dynamics over Type IIA backgrounds preserving $ \mathcal{N}=2 $ SUSY which are dual to $ \mathcal{N}=2 $ SCFTs in 4D \cite{Gaiotto:2009we}-\cite{Gaiotto:2009gz}. These solutions are typically classified in terms of a potential function\footnote{For the purpose of our present analysis, we focus only on the metric ($ g_{\mu \nu} $) and NS-NS two form ($ B_2 $) of the full $ \mathcal{N}=2 $ supersymmetric configuration.} $ V(\sigma, \eta) $ \cite{ReidEdwards:2010qs}-\cite{Nunez:2018qcj},
\begin{eqnarray}
\label{e1}
ds^2_{IIA} &=& 4f_1 (\sigma , \eta)ds^2_{AdS_5} +f_2 (\eta , \sigma)(d\sigma^2 +d\eta^2)+f_3(\eta , \sigma)d\Omega_2 (\chi , \xi)+f_4 (\sigma , \eta)d\beta^2\\
B_2 &=&f_5 (\sigma , \eta)\sin\chi d\chi \wedge d\xi\\
ds^2_{AdS_5}&=&-dt^2 \cosh^2\rho + d\rho^2 + \sinh^2\rho (d\theta^2 +\cos^2\theta d\phi^2_1 +\sin^2\theta d\phi^2_2)\\
f_1 &=&\left(\frac{2\dot{V}-\ddot{V}}{\partial^2_{\eta}V} \right)^{1/2}~,~ f_2 = f_1\frac{2\partial^2_{\eta}V}{\dot{V}}~,~f_3 = f_1\frac{2\partial^2_{\eta}V \dot{V}}{\Delta}\\
f_4 &=&f_1 \frac{4\partial^2_{\eta}V \sigma^2}{2\dot{V}-\ddot{V}}~,~f_5 = 2\left(\frac{\dot{V}\partial_{\eta}\dot{V}}{\Delta}-\eta \right),~\Delta = (2\dot{V}-\ddot{V})\partial^2_{\eta}V +(\partial_{\eta}\dot{V})^2
\end{eqnarray}
which satisfies Laplace equation with a given charge density\footnote{Here, $ \dot{V}=\sigma\partial_{\sigma}V $ and $ \ddot{V}=\sigma^2 \partial^2_{\sigma}V +\sigma \partial_{\sigma}V $.} $ \lambda (\eta) $
\begin{eqnarray}
\partial_{\sigma}(\sigma \partial_{\sigma}V)+\sigma \partial^2_{\eta}V=0~;~\lambda (\eta)=\sigma \partial_{\sigma}V(\sigma, \eta)|_{\sigma =0}.
\label{e6}
\end{eqnarray}

In order to enforce a correct quantization for charges, the charge density $ \lambda (\eta) $ must satisfy certain boundary conditions namely it must vanish at $ \eta =0 $ and must be a piece wise linear function $ \lambda = a_i \eta +q_i $ with some slope $ a_i $ \cite{ReidEdwards:2010qs}-\cite{Aharony:2012tz}. The slope changes corresponding to the position of the flavour $ D6 $ branes. On top of that, some solutions might represent SCFT quivers of finite length for which one must satisfy $ \lambda (N_c)=0 $. However there have been some exceptions where this second boundary condition is relaxed \cite{Maldacena:2000mw}.
\subsection{Spinning strings on $ AdS_5 $: A warm up}
\subsubsection{General algorithm}
We start by looking at the simplest possible \emph{folded} spinning string configuration where we turn on single spin along $ \phi_2 $ and consider the string soliton to be sitting at the equator ($ \theta =\frac{\pi}{2} $) of $ S^{3}\subset AdS_5 $. We choose to work with an ansatz where we switch off all the remaining coordinates associated to the internal spaces and only consider a factor of $ S^2 \subset S^3 $ from the $ AdS_5 $ piece of the full Type IIA solution,
\begin{eqnarray}
t = \kappa \tau ,~\rho = \rho (\tilde{\sigma}),~\phi_2 = \omega \tau ,~\sigma = \sigma (\tilde{\sigma}),~\eta = \eta (\tilde{\sigma}),~\theta = \frac{\pi}{2}~,\beta = \chi = \xi = \phi_1 =0.
\end{eqnarray}

The resulting Lagrangian density turns out to be,
\begin{eqnarray}
\mathcal{L}_P = 4\kappa^2 f_1 \cosh^2\rho -4\omega^2 f_1 \sinh^2\rho +4f_1 \rho'^2 + f_2 (\sigma'^2 +\eta'^2).
\end{eqnarray}
which yields the equations of motion of the form
\begin{eqnarray}
\label{e9}
f_1 \rho'' &=&-(\partial_{\eta}f_1 \eta' + \partial_{\sigma}f_1 \sigma' ) \rho' -(\omega^2 - \kappa^2)f_1 \cosh\rho \sinh\rho ,\\
\label{e10}
2f_2 \sigma'' &=& 4 (\kappa^2 \cosh^2\rho + \rho'^2 - \omega^2 \sinh^2\rho)\partial_{\sigma}f_1 +\partial_{\sigma}f_2(\eta'^2 - \sigma'^2 )-2\partial_{\eta}f_2 \eta' \sigma'  ,\\
2f_2 \eta'' &=& 4 (\kappa^2 \cosh^2\rho + \rho'^2 - \omega^2 \sinh^2\rho)\partial_{\eta}f_1 -\partial_{\eta}f_2 (\eta'^2 -  \sigma'^2)-2\partial_{\sigma}f_2 \eta' \sigma'.
\label{e11}
\end{eqnarray}

On the other hand, the Virasoro constraints may be schematically expressed as
\begin{eqnarray}
\label{e12}
T_{\tau \tau}&=&T_{\tilde{\sigma}\tilde{\sigma}}\nonumber\\
&=& -4\kappa^2 f_1 \cosh^2\rho + 4\omega^2 f_1 \sinh^2\rho + 4f_1 \rho'^2 +f_2 (\sigma'^2 +\eta'^2 )=0,\\
T_{\tau \tilde{\sigma}}&=&0.
\end{eqnarray}

Using (\ref{e12}) we find,
\begin{eqnarray}
\rho'^2 = \kappa^2 \cosh^2\rho -\omega^2 \sinh^2\rho -\frac{f_2}{4f_1}(\sigma'^2 + \eta'^2).
\label{e14}
\end{eqnarray}

Finally, the energy and the spin associated with the string soliton are expressed as,
\begin{eqnarray}
\label{e15}
E&=&4\frac{\sqrt{\lambda}}{4\pi}(8 \kappa)\int_0^{\rho_0}\frac{d\rho}{\rho'}f_1 \cosh^2\rho = \frac{2 \kappa\sqrt{\lambda}}{\pi}\int_0^{2\pi}d\tilde{\sigma}f_1 \cosh^2\rho\\
S&=&4\frac{\sqrt{\lambda}}{4\pi}(8 \omega)\int_0^{\rho_0}\frac{d\rho}{\rho'}f_1 \sinh^2\rho = \frac{2 \omega\sqrt{\lambda}}{\pi}\int_0^{2\pi}d\tilde{\sigma}f_1 \sinh^2\rho
\label{e16}
\end{eqnarray}
where we consider that the folded string has four segments in it where each of the segments ranges between $0\leq\rho\leq \rho_0$ such that $\rho'(\rho_0)=0$ \cite{Gubser:2002tv}.

In order to evaluate the integrals above in (\ref{e15}) and (\ref{e16}) it is therefore crucial to estimate the function $ f_1(\sigma, \eta) $ which in turn depends on the potential function $ V(\sigma, \eta) $. Therefore we start by specifying the potential function and consider its expansion close to $ \sigma =0 $ \cite{Nunez:2018qcj},
\begin{eqnarray}
V(\sigma, \eta)=\mathcal{F}(\eta)+ a \eta \log \sigma +\sum_{k=1}^{\infty}h_{k}(\eta)\sigma^{2k}
\label{e17}
\end{eqnarray}
where the coefficients $ h_{2k} $ are all fixed in terms of Laplace's equation (\ref{e6}). Notice that this is a consistent choice as $ \sigma = 
\sigma' = \sigma''=0 $ is a solution of (\ref{e10}). This also stems from the fact that $ \partial_{\sigma}f_i = 0|_{\sigma =0} $ \cite{Nunez:2018qcj}.

Below we propose some possible choices for the potential function $ V(\sigma, \eta) $.
\subsubsection{Abelian T-dual of $ AdS_5 \times S^5 $}
We start with the simplest example of Abelian T-dual of $ AdS_5 \times S^5 $ which qualitatively turns out to be the same as \cite{Gubser:2002tv}. The corresponding potential function is given by \cite{Lozano:2016kum},
\begin{eqnarray}
V_{ATD}(\sigma, \eta)=\log\sigma -\frac{\sigma^2}{2}+\eta^2 ~;~\lambda (\eta)=N_4
\label{e18}
\end{eqnarray}
where $ N_4 $ is the number of color D4 branes that create the background.

Comparing (\ref{e18}) with (\ref{e17}) we note,
\begin{eqnarray}
\mathcal{F}(\eta)= \eta^2 ~;~\eta = \frac{1}{a}=const.~;~h_1 = -\frac{1}{2}~;~h_{k>2}=0.
\label{e19}
\end{eqnarray}

Notice that substituting (\ref{e19}) into (\ref{e11}) we automatically recover the Virasoro constraint (\ref{e14}) which is an integral version of (\ref{e9}) with $ \sigma'=\eta'=0 $. Using the potential function (\ref{e18}) we finally compute the function $ f_1(\sigma, \eta) $ which eventually turns out to be $ f_1=1 $. This leads to the conserved charges of the following form,
\begin{eqnarray}
E=\frac{2 \kappa \sqrt{\lambda}}{\pi}\int_{0}^{2\pi}d\tilde{\sigma}\cosh^2\rho\\
S= \frac{2 \omega\sqrt{\lambda}}{\pi}\int_0^{2\pi}d\tilde{\sigma} \sinh^2\rho
\end{eqnarray}
which are precisely the charges obtained in the case of $ AdS_5 \times S^5 $ spinning string solutions in \cite{Gubser:2002tv}. Therefore one expects identical spectrum as that of Type IIB strings on $ AdS_5 \times S^5 $.
\subsubsection{The Sfetsos-Thompson background}
As a next example, we focus on Sfetsos-Thompson (ST) background that is obtained via non-Abelian T duality on $ AdS_5 \times S^5 $ \cite{Sfetsos:2010uq}. The corresponding potential function reads as,
\begin{eqnarray}
V_{ST}(\sigma, \eta)=\eta \log\sigma - \eta \frac{\sigma^2}{2}+\frac{\eta^3}{3}~;~\lambda (\eta)=\eta
\label{e22}
\end{eqnarray}
which represents a quiver with linearly increasing rank for the $ SU(N_c) $ color group \cite{Lozano:2016kum}.

Comparing with (\ref{e17}) we note,
\begin{eqnarray}
\mathcal{F}(\eta)= \frac{\eta^3}{3} ~;~a=1~;~h_1 = -\frac{\eta}{2}~;~h_{k>2}=0.
\end{eqnarray}

Our next task would be to compute the conserved charges associated with the string soliton. Using (\ref{e22}), it is trivial to see that both the functions $ f_1(\sigma \sim 0,\eta) $ and $ f_2(\sigma \sim 0,\eta) $ become constant and as a result the solution corresponding to (\ref{e11}) turns out to be,
\begin{eqnarray}
\eta (\tilde{\sigma})= \eta_c \tilde{\sigma}
\label{e24}
\end{eqnarray}
where $ \eta_c $ is a constant of integration. A natural interpretation of (\ref{e24}) is that the string is stretched along the $ \eta $ direction whose length is proportional to the length of the string. Locally ($ \sigma \sim 0 $) the geometry that the spinning string sees turns out to be $ AdS_5 \times R^1 $.

This finally results in the conserved charges (\ref{e15}) and (\ref{e16}) of the following form,
\begin{eqnarray}
\label{e25}
E=\frac{8\kappa\sqrt{\lambda}}{\pi}\int_{0}^{\rho_0}d\rho \frac{ \cosh^2\rho}{\sqrt{\kappa^2 \cosh^2\rho -\omega^2 \sinh^2\rho -\eta_c^2 }},\\
S=\frac{8\omega\sqrt{\lambda}}{\pi}\int_{0}^{\rho_0}d\rho \frac{ \sinh^2\rho}{\sqrt{\kappa^2 \cosh^2\rho -\omega^2 \sinh^2\rho -\eta_c^2 }}.
\label{e26}
\end{eqnarray}

Here $ \rho_0 $ corresponds to the turning point(s) which satisfies,
\begin{eqnarray}
\sinh^2\rho_0 = \frac{ \kappa^2 - \eta^2_c }{  \omega^2 -\kappa^2}
\end{eqnarray}
subjected to the fact, $ \kappa^2 > \eta^2_c $.
\subsubsection{Short strings}
Below, we evaluate the dispersion relation in two different limits. We first take into account the \emph{short} string limit namely, $ \omega \gg \sqrt{2\kappa^2 -\eta^2_c} $ which thereby yields $ \rho_0 = \frac{\sqrt{\kappa^2 -\eta^2_c}}{\sqrt{\omega^2 - \kappa^2}} $. Considering this as the primary input, a straightforward computation reveals the dispersion relation of the following form,
\begin{eqnarray}
E-S =  \frac{8 \sqrt{\lambda}}{\pi}\frac{ \kappa  \rho_0 }{\sqrt{\kappa ^2-\eta_c ^2}}= \frac{8 \sqrt{\lambda}}{\pi}\frac{\kappa}{\sqrt{\omega^2 -\kappa^2}}.
\label{e28}
\end{eqnarray}

On the other hand, the spin quantum number turns out to be,
\begin{eqnarray}
S = \frac{8 \omega \sqrt{\lambda}}{3\pi} \frac{\rho_0 ^3  }{ \sqrt{\kappa ^2-\eta_c ^2}}.
\label{e29}
\end{eqnarray}

Combining (\ref{e28}) and (\ref{e29}) we find,
\begin{eqnarray}
E -S = \sqrt{\lambda}\tilde{S}^{1/3}
\end{eqnarray}
where we define the effective spin quantum number as, $ \tilde{S}=\frac{S}{\sqrt{\lambda}}\ll 1 $. This definition is valid for both short as well as long strings.
\subsubsection{Long strings}
The second example that we consider is that of a \emph{long} string which satisfies, $ \omega \ll \sqrt{2\kappa^2 -\eta^2_c} $. This is the limit which corresponds to $ \rho_0 \gg 1 $ namely the string is considered to be stretched across the $ AdS_5 $ geometry. This leads to the dispersion relation of the following form,
\begin{eqnarray}
E-S = \frac{4\sqrt{\lambda}}{\pi}\frac{ \kappa  }{\sqrt{\kappa ^2-\eta_c ^2}}\log \Lambda
\end{eqnarray}
where $ \Lambda =  \frac{ \kappa^2 - \eta^2_c }{  \omega^2 -\kappa^2}\gg 1  $.

On the other hand, a straightforward computation on spin quantum number reveals,
\begin{eqnarray}
S = \frac{4\sqrt{\lambda}}{\pi}\frac{ \omega }{\sqrt{\kappa ^2-\eta_c ^2}} \Lambda = \sqrt{\lambda}\tilde{\beta} \Lambda ~;~ \tilde{\beta}=\frac{ 4\omega }{\pi\sqrt{\kappa ^2-\eta_c ^2}} 
\label{e32}
\end{eqnarray}

Using (\ref{e32}), we finally obtain
\begin{eqnarray}
\label{eee35}
E-S = \frac{4\sqrt{\lambda}}{\pi}\frac{ \kappa  }{\sqrt{\kappa ^2-\eta_c ^2}}\log \left( \frac{\tilde{S}}{\tilde{\beta}}\right).
\end{eqnarray}

A  precise comparison with \cite{Gubser:2002tv} reveals that our expression (\ref{eee35}) differs in two ways - firstly, we have a different constant of proportionality $ \frac{ 4\kappa  }{\sqrt{\kappa ^2-\eta_c ^2}} $ sitting in front and secondly, the log term also differs by an amount $ \log (\sqrt{\lambda}\tilde{\beta}) $.
\subsubsection{Deformed ST solution}
Let us now consider the situation in which we $ \epsilon $-deform the ST solution which therefore takes us away from the strict limit of non-Abelian T dual of $ AdS_5 \times S^5 $. This is the allowed space of solutions as the corresponding potential function \cite{Nunez:2018qcj},
\begin{eqnarray}
V_{DST}=V_{ST}+\epsilon\left( \frac{\eta^4}{12}+\frac{\sigma^4}{32}-\frac{\eta^2  \sigma^2}{4} \right). 
\label{e34}
\end{eqnarray}
still falls under the Gaiotto-Maldacena (GM) class of geometries and therefore preserves $ \mathcal{N}=2 $ SUSY \cite{Gaiotto:2009gz}. Here, $ \epsilon $ is a continuous parameter such that (\ref{e34}) is still a solution of Laplace's equation (\ref{e6}).

Comparing (\ref{e34}) with (\ref{e17}) we note down the following,
\begin{eqnarray}
\mathcal{F}(\eta)=\frac{\eta^3}{3}+\frac{\epsilon \eta^4}{12}~;~a=1~;~h_1 =-\frac{\eta}{2}-\frac{\epsilon \eta^2}{4}~;~h_2 = \frac{\epsilon}{32}~;~h_{k\geq 4}=0.
\end{eqnarray}

Using (\ref{e34}), one can further integrate (\ref{e11}) which yields,
\begin{eqnarray}
\eta'^2(\tilde{\sigma}) &=& \eta^2_c + 2 \epsilon \eta_c \int_0^{\rho_0}\frac{d\rho}{\rho'}(\omega^2 \sinh^2 \rho - \kappa^2 \cosh^2\rho) +\mathcal{O}(\epsilon^2)\nonumber\\
&=&\eta^2_c - \frac{\pi \epsilon \eta_c}{4\sqrt{\lambda} } (\kappa E - \omega S)_{ST}+\mathcal{O}(\epsilon^2).
\label{e36}
\end{eqnarray}

Finally, we note down the spectrum associated to deformed ST solution
\begin{eqnarray}
( E -  S)_{DST} =\left( 1-\frac{\epsilon \eta_c}{4}\right)(E-S)_{ST}+\cdots 
\end{eqnarray}
\subsubsection{Maldacena-Nunez solution}
The next example that we focus on is that of Maldacena-Nunez (MN) solution in Type IIA supergravity \cite{Maldacena:2000mw},
\begin{eqnarray}
\label{e38}
2 \dot{V}_{MN}&=&\sqrt{\sigma^2 +(N+\eta)^2}-\sqrt{\sigma^2 +(N-\eta)^2}\\
\lambda (\eta)&=&|N+\eta | - |N - \eta |
\end{eqnarray}
where $ N $ is the number of D4 branes that creates the background.

Using (\ref{e38}), the metric functions and their derivatives may be expanded near $ \sigma \sim 0 $
\begin{eqnarray}
\label{e40}
f_1 (\sigma \sim 0, \eta)& = & \sqrt{2} \sqrt{\sqrt{(N-\eta )^2} \sqrt{(\eta +N)^2}}\\
\partial_{\eta}f_1 (\sigma \sim 0, \eta) & = &\frac{\sqrt{2} \eta  \sqrt{\sqrt{(N-\eta )^2} \sqrt{(\eta +N)^2}}}{\eta ^2-N^2}\\
f_2 (\sigma \sim 0, \eta)& = & \frac{2 \sqrt{2}}{\sqrt{\sqrt{(N-\eta )^2} \sqrt{(\eta +N)^2}}}\\
\partial_{\eta}f_2 (\sigma \sim 0, \eta) & = &-\frac{2 \left(\sqrt{2} \eta \right)}{\sqrt{\sqrt{(N-\eta )^2} \sqrt{(\eta +N)^2}} \left(\eta ^2-N^2\right)}.
\label{e43}
\end{eqnarray}

Using (\ref{e40})-(\ref{e43}), it is easy to see that like the $ \sigma $ equation (\ref{e10}), the $ \eta $ equation (\ref{e11}) allows a trivial solution of the form $ \eta=\eta' = \eta'' = 0 $. As a result, the resulting geometry effectively boils down into an $AdS_5$ factor.  Therefore one should expect a qualitatively similar spectrum as that of \cite{Gubser:2002tv}. 

This is indeed reflected in the charges associated with the sigma model,
\begin{eqnarray}
\label{eq46}
E=\frac{8\kappa\sqrt{2N \lambda }}{\pi}\int_{0}^{\rho_0}d\rho \frac{ \cosh^2\rho}{\sqrt{\kappa^2 \cosh^2\rho -\omega^2 \sinh^2\rho  }},\\
S=\frac{8\omega\sqrt{2N\lambda}}{\pi}\int_{0}^{\rho_0}d\rho \frac{ \sinh^2\rho}{\sqrt{\kappa^2 \cosh^2\rho -\omega^2 \sinh^2\rho }}
\label{eq47}
\end{eqnarray}
which are identical to those obtained in the case of type IIB superstrings in \cite{Gubser:2002tv}.

However, there exists a non-trivial profile for $ \eta (\tilde{\sigma}) $ too. This can be obtained by using (\ref{e14}) as well as substituting (\ref{e40})-(\ref{e43}) into (\ref{e11}) which yields an equation of the form,
\begin{eqnarray}
\eta'' = 2\eta (\omega^2 \sinh^2\rho - \kappa^2 \cosh^2 \rho).
\label{e46}
\end{eqnarray}

The above equation (\ref{e46}) has a remarkably simple form,
\begin{eqnarray}
\label{eee49}
\eta'' +2\eta = 0
\end{eqnarray}
in a special situation in which we set, $ \omega = \kappa = 1 $ for simplicity. 

The most general solution can be expressed as,
\begin{eqnarray}
\eta (\tilde{\sigma})=c_2 \sin \left(\sqrt{2}\tilde{ \sigma} \right)+c_1 \cos \left(\sqrt{2} \tilde{\sigma} \right).
\end{eqnarray}

One can now combine (\ref{e15}) and (\ref{e16}) to show
\begin{eqnarray}
\label{eq52}
E-S &=&\frac{2\sqrt{\lambda}}{\pi}\int_0^{2\pi}f_1 d\tilde{\sigma}\nonumber\\
& =&\frac{2\sqrt{2\lambda}}{\pi}\int_0^{2\pi}\sqrt{N^2-\left(c_1 \cos \left(\sqrt{2} \tilde{\sigma} \right)\right){}^2} ~,
\end{eqnarray}
where we use (\ref{e40}) and set $ c_2 =0 $ without any loss of generality. 

A straightforward evaluation of the above integral (\ref{eq52}) yields a solution in terms of special Elliptic functions
\begin{eqnarray}
E-S=\frac{2\sqrt{\lambda}}{\pi }\frac{\sqrt{N^2-c_1^2 \cos ^2\left(2 \sqrt{2} \pi \right)} E\left(2 \sqrt{2} \pi |\frac{1}{1-\frac{N^2}{c_1^2}}\right)}{\sqrt{\frac{N^2-c_1^2 \cos ^2\left(2 \sqrt{2} \pi \right)}{N^2-c_1^2}}},
\end{eqnarray}
which when expanded in the large $ N $ limit reveals an expansion of the form,
\begin{eqnarray}
E-S=\frac{2\sqrt{\lambda}}{\pi }\left( 2 \sqrt{2} \pi  N-\frac{c_1^2 \left(4 \pi  \sqrt{2}+\sin \left(4 \sqrt{2} \pi \right)\right)}{8 N}\right) +\mathcal{O}(\sqrt{\lambda}/N^3) .
\end{eqnarray}

Clearly, in the strict holographic limit ($ N \rightarrow \infty $) the spectrum gets enormously simplified to yield,
\begin{eqnarray}
\label{eqd}
E-S = 4\sqrt{2}\sqrt{\lambda}N+\mathcal{O}(1/N).
\end{eqnarray}

To summarize, so far we have explored the dynamics associated with type IIA spinning strings those are mostly extended along the $ AdS_5 $ sector of the full GM geometry.  A direct comparison with \cite{Gubser:2002tv} reveals that in the short string limit, the dispersion relation for GM strings behaves like $ E \sim  S^{1/3}$ whereas for $ AdS_5 \times S^5 $ strings it behaves like $ E \sim S^{1/2} $. On the other hand, in the long string limit we come up with a logarithmic expression for the energy of the string.  However, as we have mentioned earlier, it is not quite the same as in \cite{Gubser:2002tv} due to the overall scaling factor as well as a difference term $ \log(\sqrt{\lambda}\tilde{\beta}) $.

We interpret the above result as an artifact of the stringy dynamics that is mostly confined to the $ AdS_5 $ sector of the full type IIA geometry. Being confined to the $ AdS_5 $ only, the string does not ``feel" the presence of flavor D6 branes those are located along the $ \eta $ direction of the internal manifold. On the other hand, when extended along the coordinates of the internal manifold only, these strings start seeing the presence of flavor branes which thereby modifies the corresponding dispersion relations\footnote{Following these line of arguments, for the rest of the analysis, we will therefore focus on the stringy dynamics associated to internal spaces only.}. 

Below we elaborate on this with a number of examples.
\subsection{Strings rotating on internal manifold}
Following the above discussion, we now consider rotating string configurations near the centre ($ \rho =0 $) of $ AdS_5 $ and switch on all the directions associated with the internal space. 

We propose a rotating string ansatz of the following form,
\begin{eqnarray}
t = \kappa \tau ~;~ \chi = \chi (\tilde{\sigma})~;~\sigma = \sigma (\tilde{\sigma})~;~ \eta = \eta (\tilde{\sigma})~;~\xi = \omega \tau ~;~\beta = k \tilde{\sigma}
\label{e52}
\end{eqnarray} 
where we consider the string soliton to be stretched along the polar angle ($ \chi $) of $ S^2 $ while its end points are spinning around the azimuthal direction $ \xi $ of $ S^2 $.

The corresponding Lagrangian density turns out to be,
\begin{eqnarray}
\mathcal{L}_P=4\kappa^2 f_1 - \omega^2 f_3 \sin^2\chi  + f_2 (\sigma'^2 + \eta'^2)+ f_3 \chi'^2 +k^2 f_4 -2 \omega f_5 \sin\chi \chi'.
\label{lagrangian}
\end{eqnarray}

Next, we note down equations of motion
\begin{eqnarray}
2f_2 \eta'' = \partial_{\eta}f_2 (\sigma'^2 - \eta'^2)+4 \kappa^2 \partial_{\eta}f_1 +\partial_{\eta}f_3 \chi'^2 + k^2 \partial_{\eta}f_4 \nonumber\\
-\omega^2 \partial_{\eta}f_3 \sin^2\chi -2\omega \partial_{\eta}f_5 \sin\chi \chi' -2\partial_{\sigma}f_2 \eta' \sigma',
\label{e53}
\end{eqnarray}
\begin{eqnarray}
2f_2 \sigma'' = \partial_{\sigma}f_2 ( \eta'^2 - \sigma'^2 )+4 \kappa^2 \partial_{\sigma}f_1 +\partial_{\sigma}f_3 \chi'^2 + k^2 \partial_{\sigma}f_4 \nonumber\\
-\omega^2 \partial_{\sigma}f_3 \sin^2\chi -2\omega \partial_{\sigma}f_5 \sin\chi \chi' -2\partial_{\eta}f_2 \eta' \sigma',
\label{e54}
\end{eqnarray}
\begin{eqnarray}
\label{chi}
f_3 \chi'' = \omega (\partial_{\eta}f_5 \eta' + \partial_{\sigma}f_5 \sigma')\sin\chi -(\partial_{\eta}f_3 \eta' + \partial_{\sigma}f_3 \sigma')\chi' - \omega^2 f_3 \sin\chi \cos\chi,
\end{eqnarray}
those are supplemented by the Virasoro constraints of the following form
\begin{eqnarray}
\label{e56}
T_{\tau \tau}&=&T_{\tilde{\sigma}\tilde{\sigma}}\nonumber\\
&=& -4\kappa^2 f_1 +\omega^2 f_3 \sin^2\chi +f_2 (\sigma'^2 +\eta'^2 )+f_3 \chi'^2 +k^2 f_4=0,\\
T_{\tau \tilde{\sigma}}&=&0.
\end{eqnarray}

Finally, we note down the energy and the R-charge associated with the sigma model
\begin{eqnarray}
E&=&\frac{8 \sqrt{\lambda}\kappa}{\pi}\int_{0}^{\chi_0}\frac{d \chi}{\chi'}f_1 =\frac{2 \sqrt{\lambda}\kappa}{\pi}\int_{0}^{2 \pi} d\tilde{\sigma}f_1,\\
J &=& \frac{2 \sqrt{\lambda}\omega}{\pi}\int_{0}^{\chi_0}\frac{d \chi}{\chi'}f_3 \sin^2\chi = \frac{ \sqrt{\lambda}\omega}{2\pi}\int_{0}^{2 \pi}d\tilde{\sigma}f_3 \sin^2\chi.
\end{eqnarray}

Like before, we see that $ \sigma =\sigma' =\sigma'' =0 $ is a solution of (\ref{e54}). Therefore, the expansion (\ref{e17}) of the potential function near $ \sigma =0 $ is still valid. Considering these facts, below we focus on some interesting cases.
\subsubsection{The Sfetsos-Thompson background}
Given the ST potential (\ref{e22}), the corresponding $ \eta $ equation (\ref{e53}) takes the following form
\begin{eqnarray}
\eta'' = \frac{\eta (\chi'^2 -\omega^2 \sin^2\chi) }{\left(4 \eta ^2+1\right)^2}+\frac{2\omega \eta ^2 \left(4 \eta ^2+3\right)}{\left(4 \eta ^2+1\right)^2}\sin\chi \chi'.
\label{e60}
\end{eqnarray}

On the other hand, the constraint (\ref{e56}) yields
\begin{eqnarray}
\chi'^2 = \frac{1}{\eta^2}(\kappa^2 -\eta'^2 )\left(  4\eta^2 +1\right) -\omega^2 \sin^2\chi.
\label{e61}
\end{eqnarray}

Finding an exact analytic solution of (\ref{e60}) is indeed a quite nontrivial task. However, a careful look reveals that asymptotic solutions are indeed available. For example, we propose a solution of the form, $ \chi' (\tilde{\sigma}) =  \omega \sin\chi $ in the region $ \eta \sim 0 $ of the $ \mathcal{N}=2 $ quiver. 

A consistent solution satisfying both (\ref{e60}) and (\ref{e61}) turns out to be,
\begin{eqnarray}
\eta (\tilde{\sigma})& = & \kappa \tilde{\sigma}~;~|\tilde{\sigma}| \ll 1.
\end{eqnarray} 

On the other hand, in the domain $ \eta \gg 1 $ the corresponding solutions turn out to be
\begin{eqnarray}
\eta (\tilde{\sigma}) & = & -2 \omega \int_0^{\chi_0}\frac{d\chi}{\chi'}\cos\chi,\\
\chi' (\tilde{\sigma})& = & \sqrt{| 4\kappa^2 -\omega^2 |}.
\end{eqnarray}

Using the above set of solutions (both for the large and the small $ \eta $ regime), below we note down the energy as well as the R-charge associated with the string configuration
\begin{eqnarray}
\label{eqE1}
E & = & \frac{8 \sqrt{\lambda}\kappa}{\pi}\left(\omega^{-1}\int_{\eta \sim 0}\frac{d\chi}{\sin\chi}+\int_{\eta \gg 1} \frac{d \chi}{\sqrt{|4\kappa^2 - \omega^2 |}}+\int_{0<\eta <\infty}\frac{d \chi}{\chi'}\right)\nonumber\\
& = & \frac{8 \sqrt{\lambda}\kappa}{\pi} \left(\kappa^{-1} \int_{\eta \sim 0}d\eta +\frac{\pi}{2\sqrt{|4\kappa^2 -\omega^2 |}}+\int_{0<\eta <\infty}\frac{d \chi}{\chi'} \right),
\end{eqnarray}
\begin{eqnarray}
\label{eqJ1}
J  &=& \frac{2\sqrt{\lambda}}{\pi}\left(1-\frac{\omega}{4}\int_{\eta \sim 0}d\tilde{\sigma}\frac{\sin^2\chi}{1+4\kappa^2 \tilde{\sigma}^2}+\frac{\pi \omega}{4\sqrt{|4\kappa^2 -\omega^2 |}} +\omega \int_{0<\eta <\infty}\frac{d\chi}{\chi'}\frac{4\eta^2 \sin^2\chi}{4\eta^2 +1}\right),\nonumber\\
&=& \frac{2\sqrt{\lambda}}{\pi}\left(\frac{3}{4}+\frac{\pi \omega}{4\sqrt{|4\kappa^2 -\omega^2 |}} +\omega \int_{0<\eta <\infty}\frac{d\chi}{\chi'}\frac{4\eta^2 \sin^2\chi}{4\eta^2 +1} \right). 
\end{eqnarray}

Using (\ref{eqE1}) and (\ref{eqJ1}), we finally arrive at the dispersion relation of the following form
\begin{eqnarray}
E-J &=& \frac{7\sqrt{\lambda}}{2\sqrt{3}}\left(1-\frac{3\sqrt{3}}{7\pi}+\frac{16 \sqrt{3}}{7 \pi}\mathcal{I} \right),\\
\mathcal{I}&=&\int_{0<\eta <\infty}\frac{d\chi}{\chi'}\left(1-\frac{\eta^2 \sin^2 \chi}{4\eta^2 +1} \right),
\label{eqI}
\end{eqnarray}
where for simplicity we set, $ \omega = \kappa =1 $. Clearly, an analytic evaluation of the entity (\ref{eqI}) is indeed a challenging task as this amounts of solving the set of equations (\ref{e60}) and (\ref{e61}) \emph{exactly} for an intermediate value of $ \eta $. However, a numerical evaluation of these equations (and therefore $ \mathcal{I} $) may be carried out by choosing appropriate initial conditions for $ \eta $ and $ \chi $ (and their derivatives) such that the Virasoro constraint (\ref{e56}) is satisfied.

\subsubsection{Maldacena-Nunez solution}
The next example that we consider is that of MN solution as depicted in (\ref{e38}). As usual, the $ \sigma $-equation (\ref{e54}) is satisfied by setting $ \sigma =\sigma'=\sigma'' =0 $. However, on the other hand, the exact solution of the $ \eta $-equation (\ref{e53}) is indeed quite nontrivial. 

In the holographic limit ($ N \gg 1 $), the conserved charges associated with the MN background turns out to be
\begin{eqnarray}
E & = & \frac{2 \sqrt{\lambda}\kappa}{\pi}\left(2 \sqrt{2}\pi N -\frac{1}{\sqrt{2} N}\int_0^{2\pi}d\tilde{\sigma}\eta^2 +\cdots \right), \\
J & = &\frac{\sqrt{\lambda}\omega}{2 \pi}\left(\frac{2 \sqrt{2} }{N}\int_{0}^{2\pi}d\tilde{\sigma}\eta^2 \sin^2\chi + \cdots \right). 
\end{eqnarray}

Finally, setting $ \omega = \kappa $ leads to a dispersion relation of the form,
\begin{eqnarray}
E -J = \sqrt{\lambda}N - \frac{\sqrt{2}\sqrt{\lambda}}{\pi N}\int_{0}^{2\pi}d\tilde{\sigma}\eta^2 (1+\sin^2\chi)+\cdots
\end{eqnarray}
Clearly, in the strict large N ($ \rightarrow \infty $) limit we obtain a dispersion relation, $ E-J =\sqrt{\lambda}N$ similar to that of (\ref{eqd}).
\subsubsection{Remarks on general GM geometries}
We now generalize our calculations for the generic GM potential of the form \cite{ReidEdwards:2010qs}-\cite{Aharony:2012tz}, \cite{Nunez:2019gbg}
\begin{eqnarray}
V (\sigma , \eta) = -\sum_{n=1}^{\infty}\frac{c_n}{w_n}K_0  (w_n \sigma)\sin (w_n \eta)~;~w_n = \frac{n\pi}{N_5}
\label{e71}
\end{eqnarray}
which is defined for the entire range of the $ \sigma $ direction.

Using (\ref{e71}), the potential functions $ f_i (\sigma, \eta) $ may be estimated in a straightforward way near $ \sigma \sim 0 $. This leads to the conserved charges of the following form
\begin{eqnarray}
E&= & \frac{2\sqrt{\lambda}\kappa}{\pi}\int_{0}^{2\pi}d\tilde{\sigma}\frac{\sqrt{\sum_{n=1}^{\infty}\frac{2 c_n \sin \left(\eta  w_n\right)}{w_n}}}{\sqrt{\sum_{n=1}^{\infty}c_n w_n \sin \left(\eta  w_n\right) \left(\log \left(\frac{\sigma w_n}{2}\right)+\gamma_{E} \right)}},\\
J &= & \frac{\sqrt{\lambda}\omega}{\pi}\int_0^{2\pi}d\tilde{\sigma}\frac{f_1 \sin^2\chi \sum_{n=1}^{\infty}c^2_n  \sin^2 \left(\eta  w_n\right) \left(\log \left(\frac{\sigma w_n}{2}\right)+\gamma_{E} \right)}{\sum_{n=1}^{\infty}c_n^2 \left(\cos ^2\left(\eta  w_n\right)-2 \sin ^2\left(\eta  w_n\right) \left(\log \left(\frac{\sigma w_n }{2}\right)+\gamma_{E} \right)\right)}
\end{eqnarray}
where $\gamma_E$ is the Euler Gamma function. Clearly, the exact evaluation of these integrals is indeed quite difficult in the small $ \sigma $ limit. 

Likewise, considering $\sigma \gg 1$ we obtain the charges of the following form,
\begin{eqnarray}
\label{E}
E&= & \frac{2\sqrt{\lambda}\kappa}{\pi}\int_{0}^{2\pi}d\tilde{\sigma}\sigma (\tilde{\sigma}) \\
J &= & \frac{\sqrt{\lambda}\omega}{\pi}\int_0^{2\pi}d\tilde{\sigma}\frac{\sin^2\chi \sum_{n=1}^{\infty}e^{-2 \sigma  w_n} \sin ^2\left(\eta  w_n\right) c_n^2}{\sum_{n=1}^{\infty}e^{-2 \sigma  w_n}  c_n^2 w_n}\nonumber\\
& = &\frac{\sqrt{\lambda}\omega N_5}{\pi^2}\int_0^{2\pi}d\tilde{\sigma} \sin^2\chi \sin^2 (\frac{\eta \pi}{N_5})+\cdots \nonumber\\
& = &\frac{\sqrt{\lambda}\omega }{N_5}\int_0^{2\pi}d \tilde{\sigma}\eta^2 \sin^2\chi + \mathcal{O}(1/N^3_5)
\label{J}
\end{eqnarray}
where we suppress all the sub leading terms being very small in the large $ N_5 $ limit. 

Combining (\ref{E}) and (\ref{J}) we extract a dispersion relation of the form, 
\begin{eqnarray}
\label{d}
E-J \Big |_{\sigma \gg 1}=\frac{\sqrt{\lambda}}{\pi}\int_0^{2\pi} d\tilde{\sigma}\left( 2 \sigma - \frac{\pi}{N_5}\eta^2 \sin^2\chi \right) ~;~\kappa = \omega =1
\end{eqnarray}
where the exact analytic evaluation of the integral in (\ref{d}) is indeed challenging. However, a numerical evaluation is possible where one solves the dynamics (\ref{e53})-(\ref{chi}) by choosing appropriate initial conditions for the variables $ \eta $ , $ \sigma $ and $ \chi $ (and their first derivatives) such that the Virasoro constraint (\ref{e56}) is satisfied.
\subsubsection{Strings passing through flavor branes}
\label{strings through flavor branes}
We now focus on spinning string states those are stretched through the flavor $ D6 $ branes located along the $ \eta $ direction of the internal manifold. We start with the previous ansatz (\ref{e52}) for the rotating string except for the fact that now we set, $ \chi = \frac{\pi}{2} $. In other words, we consider the string soliton to be sitting at the equatorial plane of $ S^2 $.

This leads to the equations of motion of the following form,
\begin{eqnarray}
2f_2 \eta'' = \partial_{\eta}f_2 (\sigma'^2 - \eta'^2)+4 \kappa^2 \partial_{\eta}f_1 + k^2 \partial_{\eta}f_4 
-\omega^2 \partial_{\eta}f_3 -2\partial_{\sigma}f_2 \eta' \sigma',
\label{EE76}
\end{eqnarray}
\begin{eqnarray}
2f_2 \sigma'' = \partial_{\sigma}f_2 ( \eta'^2 - \sigma'^2 )+4 \kappa^2 \partial_{\sigma}f_1 + k^2 \partial_{\sigma}f_4 
-\omega^2 \partial_{\sigma}f_3  -2\partial_{\eta}f_2 \eta' \sigma'.
\label{EE77}
\end{eqnarray}

The corresponding Virasoro constraints are given by,
\begin{eqnarray}
\label{EE78}
T_{\tau \tau}&=&T_{\tilde{\sigma}\tilde{\sigma}}\nonumber\\
&=& -4\kappa^2 f_1 +\omega^2 f_3  +f_2 (\sigma'^2 +\eta'^2 ) +k^2 f_4=0,\\
T_{\tau \tilde{\sigma}}&=&0.
\end{eqnarray}

Finally, we note down the conserved charges associated with the sigma model
\begin{eqnarray}
E &=&\frac{4 \kappa \sqrt{\lambda}}{\pi}\int_{0}^{\eta_{max}}f_1 \frac{d\eta}{\eta'},\\
J &=&\frac{\omega \sqrt{\lambda}}{\pi}\int_0^{\eta_{max}}f_3 \frac{d\eta}{\eta'}
\label{EE81}
\end{eqnarray}
where the picture that we have in mind is that of a folded string that is made up of two segments where each segment is stretched between $ [0, \eta_{max}] $.

The above configuration turns out to be an interesting situation to look for and in particular to explore what happens to the spectrum as the string passes through $ N_6 $ flavor D6 branes located along the $ \eta $ direction. 

We address these issues considering the following two examples:\\\\
$ \bullet ~ \textbf{Single kink:} $ The first example we consider is that of a \emph{single kink} solution \cite{Lozano:2016kum} where one considers $ N_6 $ flavor D6 branes sitting at $ \eta =P $ of the internal manifold.  

The most general expression of the charge density is given by \cite{Nunez:2018qcj},
\begin{eqnarray}
\dot{V}_k(\eta , \sigma)=\frac{N_6}{2}\sum_{n=-\infty}^{\infty}(P+1)[\sqrt{\sigma^2 +(\eta +P -2n(1+P))^2}-\sqrt{(\eta -2 n (P+1)-P)^2+\sigma ^2}]\nonumber\\
+P[\sqrt{(\eta -2 n (P+1)-P-1)^2+\sigma ^2}-\sqrt{(\eta -2 n (P+1)+P+1)^2+\sigma ^2}].
\label{E76}
\end{eqnarray}

An expansion of (\ref{E76}) near $ \sigma \sim 0 $ reveals the corresponding potential function as,
\begin{eqnarray}
V_k (\sigma \sim 0, \eta)= \eta N_6 \log\sigma +\frac{\eta N_6 \sigma^2}{4}\Lambda_{k}(\eta , P)
-\frac{\eta N_6 \sigma^2}{4}\frac{(P+1)}{(P^2 -\eta^2)}
\label{E53}
\end{eqnarray}
where we denote the function\footnote{For the purpose of the present paper, we restrict ourselves to the range $ -k\leq n \leq  k$ where $ k $ can be set as large as possible.},
\begin{eqnarray}
\Lambda_{k}(\eta , P) =(P+1)\sum_{m=1}^{k}\frac{1}{(2m+(2m-1)P)^2-\eta^2}-\frac{1}{(2m+(2m+1)P)^2-\eta^2}\nonumber\\
+\frac{P}{(2k+1)^2(1 +P)^2 -\eta^2}.
\end{eqnarray}

Using (\ref{E53}), the corresponding potential functions may be enumerated as below
\begin{eqnarray}
\label{EE85}
f_1(\sigma \sim 0, \eta)&= & 4f^{-1}_2(\sigma \sim 0, \eta),\\
\label{EE86}
f_2(\sigma \sim 0, \eta)&= &\frac{\sqrt{2} }{\sqrt{\frac{\eta  \left(P^2-\eta ^2\right)^3}{\sigma ^2 \left(\left(P^2-\eta ^2\right)^3 \left(\eta  \partial^2_{\eta}\Lambda (\eta , P )+2 \partial_{\eta}\Lambda (\eta , P )\right)-2 \eta  (P+1) \left(\eta ^2+3 P^2\right)\right)}}},\\
f_3(\sigma \sim 0, \eta)&= &  \eta^2 f_2(\sigma \sim 0, \eta),\\
f_4(\sigma \sim 0, \eta)&= &\sigma^2 f_2(\sigma \sim 0, \eta).
\end{eqnarray}

Unlike the previous examples, we see that $ \sigma =\sigma' = \sigma''=0 $ is not a solution to the $ \sigma $ equation (\ref{EE77}). On the orher hand, a careful analysis of (\ref{EE77}) near $ \sigma \sim 0 $ reveals,
\begin{eqnarray}
\eta'^2 -\sigma'^2 = \omega^2 \eta^2 +\frac{16 \kappa^2}{f^2_2}
\end{eqnarray}
which together with (\ref{EE78}) yields,
\begin{eqnarray}
\eta' = \frac{4\kappa}{f_2}.
\label{E91}
\end{eqnarray}

Using (\ref{E91}), the conserved charges associated with the sigma model turns out to be
\begin{eqnarray}
\label{EE92}
E_k & = & \frac{4\sqrt{\lambda}}{\pi}(P+1),\\
J_k & = & \frac{\omega \sqrt{\lambda}}{4 \pi \kappa }(P+1)^2 \sigma_c^2 \partial_{\eta}\Lambda_{k} (P+1,P)-\frac{(P+1)\omega \sqrt{\lambda}\sigma^2_c}{2\pi \kappa}\int_0^{P+1}\frac{\eta^2(\eta^2 +3P^2)}{(P^2 -\eta^2)^3}
\label{EE93}
\end{eqnarray}
where we fix the string at $ \sigma \sim \sigma_c \ll 1$ as the final value of the integral (\ref{EE81}) does not depend on the location of the string along the $ \sigma $ axis.

Clearly, the energy ($ E_k $) of the string is directly proportional to the length of the quiver. On the other hand, the expression for the R-charge ($ J_k $) carries the $ \eta $ derivative of the function $ \Lambda_{k}(\eta, P) $ which therefore encodes the information about the flavor D6 branes those are located along the $ \eta $ axis. However, the nontrivial contribution due to flavor branes comes from the second term in (\ref{EE93}) where the integral clearly exhibits a singularity at the location ($ \eta = P $) of the D6 branes. The key source to this singularity could in fact be traced back into the preexisting singularity in the potential function (\ref{E53}) and thereby in the metric function (\ref{EE86}) at the location ($ \eta =P $) of these flavor D6 branes.

In order to take care of the pole structure around $ \eta \sim P $, we divide the entire range of the $ \eta $ integral (\ref{EE93}) into two sub-intervals namely, $ [0,P-\epsilon] $ and $ [P+\epsilon, P+1] $ where one has to consider the limit $ \epsilon \rightarrow 0 $ towards the end of the calculation. 

Combining all these pieces together we find,
\begin{eqnarray}
\int_0^{P+1}\frac{\eta^2(\eta^2 +3P^2)}{(P^2 -\eta^2)^3}= \frac{(P+1)^3}{(2 P+1)^2}-\frac{1}{\epsilon }
\label{EEE93}
\end{eqnarray}
where $ 1/\epsilon $ divergence arises due to some short distance effects as a result of strings ending on flavor branes. In order to deal with such divergences one needs to introduce suitable counter terms that lead to what we identify as \emph{regularized} R-charge ($ J_k^{(reg)} = J_k+\frac{1}{\epsilon}$). For simplicity, we will not use the superscript anymore.

On the other hand, a closer look reveals that for any generic value of $ k $ one obtains,
\begin{eqnarray}
\partial_{\eta}\Lambda_{k} (P+1,P)=2c(k)(1+P)^2 +\frac{P}{8 k^2\left(k+1\right)^2 (P+1)^3}
\label{EEE94}
\end{eqnarray}
where the coefficient $ c(k) $ represents an infinite series of the form,
\begin{eqnarray}
c(k)=\sum_{m=1}^{k}\frac{1}{\left((P+1)^2-(P-2 m (P+1))^2\right)^2}-\frac{1}{(2 P+1-4 m (P+1) (m P+m+P))^2}.
\end{eqnarray}

Combining (\ref{EEE93}) and (\ref{EEE94}) one can finally express the regularized R-charge as,
\begin{eqnarray}
\frac{J_k}{\sigma^2_c} = \frac{\omega \sqrt{\lambda}}{2\pi \kappa}(P+1)^4\left(c(k)-\frac{1}{(2P+1)^2} \right)
\label{EEE96}
\end{eqnarray}
where we set the upper limit of the summation as $ k\rightarrow \infty $ while keeping $ P $ fixed.

Finally, using (\ref{EE92}), it is now straightforward to invert (\ref{EEE96}) which yields a dispersion relation of the form,
\begin{eqnarray}
E_k = \beta_c (k) \lambda^{3/8}J_k^{1/4}
\label{closed string kink}
\end{eqnarray}
where we define the function,
\begin{eqnarray}
\beta_c (k) =  \frac{2^{9/4}\kappa^{1/4}}{\omega^{1/4}\pi^{3/4}\sigma_c^{1/2}}\left(c(k)-\frac{1}{(2P+1)^2} \right)^{-\frac{1}{4}}. 
\end{eqnarray}\\
$ \bullet ~ \textbf{Uluru profile:} $ The second background we consider goes under the name of Uluru profile \cite{ReidEdwards:2010qs}. The corresponding charge density is given by \cite{Lozano:2016kum}
\begin{eqnarray}
\dot{V}_{u}(\sigma, \eta)=\frac{N_6}{2}\sum_{m=-\infty}^{\infty}\sum_{l=1}^{3}\sqrt{\sigma^2 +(\nu_l +2m(2P+K)-\eta)^2}\nonumber\\
-\sqrt{\sigma^2 +(\nu_l -2m(2P+K)+\eta)^2}
\label{EE96}
\end{eqnarray}
together with the coefficients, $ \nu_1 = P $, $ \nu_2 = P+K $ and $ \nu_3=-(2P+K) $. The above background corresponds to putting $ N_6 $ flavor D6 branes at $ \eta =P $ and $ \eta =P+K $ along the $ \eta $ axis of the internal manifold.

Like before, we expand (\ref{EE96}) around $ \sigma \sim 0 $ which yields the potential function of the following form,
\begin{eqnarray}
V_{u}(\sigma \sim 0, \eta)= -\eta N_6 \log\sigma +\frac{\eta N_6 \sigma^2}{4(P^2 -\eta^2)}+\frac{\eta N_6 \sigma^2}{4}\Lambda_{u}(\eta , K, P)
\label{EE98}
\end{eqnarray}
where we denote,
\begin{eqnarray}
\Lambda_{u}(\eta , K, P)=\sum_{n=1}^{u}(-1)^{n+1}\left(\frac{1}{(n K+(2n-1)P)^2-\eta^2} -\frac{1}{(n K+(2n+1)P)^2-\eta^2}\right)
\end{eqnarray}
where for the present case we restrict ourselves to the bound, $ -u \leq m \leq u $.

Using (\ref{EE98}), the corresponding metric functions turn out to be
\begin{eqnarray}
\label{EE100}
f_1(\sigma \sim 0, \eta)&= &4 f^{-1}_2(\sigma \sim 0, \eta),\\
f_2(\sigma \sim 0, \eta)&= & \frac{\sqrt{2} }{ \sqrt{\frac{\eta  \left(\eta ^2-P^2\right)^3}{\sigma ^2 \left(2 \eta ^3+\left(P^2-\eta ^2\right)^3 \left(\eta  \partial^2_{\eta}\Lambda_u (\eta , K, P)+2 \partial_{\eta}\Lambda_u (\eta , K, P)\right)+6 \eta  P^2\right)}}},\\
f_3(\sigma \sim 0, \eta)&= &\eta^2 f_2(\sigma \sim 0, \eta),\\
f_4(\sigma \sim 0, \eta)&= &\sigma^2 f_2(\sigma \sim 0, \eta).
\label{EE103}
\end{eqnarray}

Qualitatively, one therefore has the similar situation as in the case of one kink solution mentioned above. This leads to an identical expression for $ \eta' $ as the structure of the $ \sigma $ equation (\ref{EE77}) does not alter.  

Finally, using the above results (\ref{EE100})-(\ref{EE103}), the energy ($ E_u $) and the R- charge ($ J _u$) associated with the sigma model turns out to be,
\begin{eqnarray}
\label{EE104}
E_u &=&\frac{4\sqrt{\lambda}}{\pi}(2P+K)\\
J_u&=&\frac{\omega \sqrt{\lambda}\sigma^2_c}{2\pi \kappa}(2P+K)^2 \partial_{\eta}\Lambda_{u}(2P+K, K, P)+\frac{\omega \sqrt{\lambda}\sigma^2_c}{\pi \kappa}\int_{0}^{2P+K}\frac{\eta^2(\eta^2 +3P^2)}{( P^2-\eta^2)^3}.
\label{EE105}
\end{eqnarray}

Like in the previous example of single kink solution, the nontrivial contribution due to the presence of the flavor branes shows up in the expression for the R-charge (\ref{EE105}) which diverges precisely at the location ($ \eta =P $) of the flavor D6 branes. 

Following our previous approach, we divide the integral in (\ref{EE105}) into three segments namely, $ [0,P-\epsilon] $,  $ [P+\epsilon, P+K] $ and $ [P+K, 2P+K] $  where $ \eta =  P $ and $ \eta = P+K $ are precisely the locations of the D6 branes along the $ \eta $ axis. This clearly separates out the divergent piece around $ \eta \sim P $ which finally yields,
\begin{eqnarray}
\int_{0}^{2P+K}\frac{\eta^2(\eta^2 +3P^2)}{( P^2 -\eta^2)^3}= \frac{(K+2 P)^3}{(K+P)^2 (K+3 P)^2}-\frac{1}{\epsilon }
\end{eqnarray}
where $ 1/\epsilon $ divergence is clearly an artifact of strings ending on D6 branes at $ \eta =P $. 

Like before, the above divergence can be taken care of by incorporating suitable counter term that leads to finite R-charge for the $ \mathcal{N}=2 $ SCFTs,
\begin{eqnarray}
J_u = \frac{\omega \sqrt{\lambda}\sigma^2_c}{2\pi \kappa}(2P+K)^2 \partial_{\eta}\Lambda_{u}(2P+K, K, P)+\frac{\omega \sqrt{\lambda}\sigma^2_c}{\pi \kappa}\frac{(K+2 P)^3}{(K+P)^2 (K+3 P)^2}.
\label{EE109}
\end{eqnarray}

A straightforward computation further reveals,
\begin{eqnarray}
\partial_{\eta}\Lambda_{u}(2P+K, K, P)=2(2P+K)c(u)
\end{eqnarray}
where define the series,
\begin{eqnarray}
c(u)=\sum_{n=1}^{u} \frac{ (-1)^{n+1} }{\left((K+2 P)^2-(K n+(2 n-1) P)^2\right)^2}\nonumber\\
-\frac{ (-1)^{n+1} }{\left((K+2 P)^2-(K n+2 n P+P)^2\right)^2}.
\end{eqnarray}

Using (\ref{EE109}), it is now trivial to see,
\begin{eqnarray}
\frac{J_u}{\sigma^2_c} = \frac{\omega \pi^2 E^3_u}{2^6 \kappa \lambda}\left(c(u)+\frac{1}{(K+P)^2(K+3P)^2} \right) = \frac{\gamma_c^3(u)}{\lambda}E^3_u
\end{eqnarray}
which therefore leads to a dispersion relation of the form,
\begin{eqnarray}
E_u = \frac{\gamma_c^{-1} (u)}{\sigma^{2/3}_c}\lambda^{1/3}J_u^{1/3}.
\end{eqnarray}\\
$ \bullet $ \textbf{An interpretation of the $ 1/\epsilon $ divergence:} Below we provide a \emph{geometrical} interpretation of the divergences appearing in both the R- charge(s) as the string approaches the flavor D6 branes. As mentioned above, the key source to these divergences is hidden in the geometrical singularity caused due to flavor D6 branes at $ \eta =P $. Therefore, it takes infinite amount of energy for the string to actually reach D6 branes. In case of rotating strings, this corresponds to infinite rotational energy ($ \sim L^2 $) which thereby corresponds to an infinite angular momentum to reach the flavor branes. From this perspective, $ \epsilon $ plays the role of an UV cutoff on the (rotational) energy of the string. From the geometrical point of view, this corresponds to the closest approach of the string to the stack of flavor D6 branes located at $ \eta =P $.

A nice interpretation of this comes from the effective string Lagrangian description of rotating strings as depicted above in (\ref{lagrangian}),
\begin{eqnarray}
\mathcal{L}_P=f_2(\sigma'^2 + \eta'^2)+V_{eff}
\end{eqnarray}
where the \emph{effective} potetial that the string sees turns out to be,
\begin{eqnarray}
V_{eff}=4\kappa^2 f_1 -\omega^2 f_3 +k^2 f_4.
\label{veff}
\end{eqnarray}

Taking the specific example of single kink profile, below we explore the potential function (\ref{veff}) near the singularity $ \eta \sim P $. However, similar conclusion can also be drawn for the Uluru profile which we prefer not to repeat here.

Expanding (\ref{veff}) for the single kink profile we find,
\begin{eqnarray}
V^{(k)}_{eff}(\sigma \sim 0, \eta)= \frac{16 \kappa^2}{f_2}-\omega^2 \eta^2 f_2.
\label{potential}
\end{eqnarray}

A careful analysis reveals that near $ \eta \sim P $ the potential function (\ref{potential}) has a pole of the form,
\begin{eqnarray}
V^{(k)}_{eff}(\sigma \sim 0, \eta \sim P)= -\frac{2\sigma_c\omega^2 P^2 (P+1)^{1/2}}{| \epsilon|^{3/2}}
\label{eee117}
\end{eqnarray}
which clearly shows an infinite deep in the effective potential as the string approaches the flavor branes. Therefore strings eventually do not reach the flavor branes.  As a result, one could imagine a mixed configuration that is a collection of rotating folded strings floating around D6 branes at $ \eta \sim P $ (with energy (\ref{closed string kink})) together with open strings that live on world-volume of flavor D6 branes.
\subsection{Remarks on folded string solutions without spin}
We now consider non rotating folded string configurations those are extended along the $ \eta $ direction  of the internal space as well as wrapping the isometry ($ \xi $) of $ S^2 $. 

To this end, we propose a different ansatz of the following form,
\begin{eqnarray}
t = \kappa \tau ~;~ \chi = \chi (\tilde{\sigma})~;~\sigma = \sigma (\tilde{\sigma})~;~ \eta = \eta (\tilde{\sigma})~;~\xi = l \tilde{\sigma} ~;~\beta = k \tilde{\sigma}
\end{eqnarray}
where we consider the string soliton to be sitting at the center of $ AdS_5 $ while simultaneously wrapping all the isometry direction of the internal manifold. As we proceed further, we see that the winding number $ l $ of the string plays a vital role to the string spectrum as the string passes through flavor branes along the $ \eta $ direction.

The corresponding Lagrangian density turns out to be,
\begin{eqnarray}
\mathcal{L}_P = 4 \kappa^2 f_1 +f_2 (\sigma'^2 + \eta'^2)+f_3 (\chi'^2 +l^2  \sin^2\chi) +k^2 f_4.
\label{EE119}
\end{eqnarray}

The resulting equations of motion are given by
\begin{eqnarray}
\label{e78}
2f_2 \eta'' &=&4\kappa^2 \partial_{\eta}f_1 +\partial_{\eta}f_2 (\sigma'^2 -\eta'^2)+\partial_{\eta}f_3 (\chi'^2 + l^2 \sin^2\chi )+k^2\partial_{\eta}f_4 -2\partial_{\sigma}f_2 \sigma'\eta', \\
2f_2 \sigma'' &=&4\kappa^2 \partial_{\sigma}f_1 +\partial_{\sigma}f_2 (\eta'^2 - \sigma'^2)+\partial_{\sigma}f_3 (\chi'^2 + l^2 \sin^2\chi )+k^2\partial_{\sigma}f_4-2\partial_{\eta}f_2 \sigma'\eta', \\
\label{e79}
f_3 \chi'' &=& -(\partial_{\eta}f_3 \eta' + \partial_{\sigma}f_3 \sigma')\chi' +l^2 f_3 \sin\chi \cos\chi,
\label{e80}
\end{eqnarray}
those are supplemented by the Virasoro constraints of the form 
\begin{eqnarray}
\label{e81}
T_{\tau \tau}&=&T_{\tilde{\sigma}\tilde{\sigma}}\nonumber\\
&=& -4\kappa^2 f_1 +f_2 (\sigma'^2 +\eta'^2 )+f_3 (\chi'^2 +l^2 \sin^2\chi)+k^2 f_4=0,\\
T_{\tau \tilde{\sigma}}&=&0.
\end{eqnarray}
\subsubsection{Sfetsos-Thompson and its deformation}
We start with the example of ST solution (\ref{e22}) expanded near $ \sigma =0 $. This yields the energy associated with these extended string configurations as,
\begin{eqnarray}
E_{ST}=\frac{4\sqrt{\lambda}\kappa}{\pi}\int_0^{\eta_{max}}\frac{d\eta}{\eta'}.
\end{eqnarray}

Like before, the $ \sigma $-equation (\ref{e79}) is trivially satisfied by setting $ \sigma''=\sigma'=\sigma =0 $. On the other hand, the $ \eta $- equation (\ref{e78}) yields,
\begin{eqnarray}
\eta'' = \frac{\eta }{\left(4 \eta ^2+1\right)^2}(\chi'^2 +l^2 \sin^2\chi).
\label{e84}
\end{eqnarray}

Using (\ref{e81}), we can simplify (\ref{e84}) further to yield
\begin{eqnarray}
\eta '' =\frac{(\kappa^2 - \eta'^2)}{\eta (1+4 \eta^2)}.
\label{e85}
\end{eqnarray}

Integrating (\ref{e85}) and setting $ \eta_{max}\gg 1 $ we find the energy of the extended string as,
\begin{eqnarray}
E_{ST} = \alpha_c \sqrt{\lambda} ~\eta_{max}+\cdots
\label{e86}
\end{eqnarray}
where, $ \alpha_c = \frac{4\kappa}{\pi\sqrt{4 e^{2 c_1}+\kappa ^2}} $ is a constant together with $ c_1 $ being the constant of integration that could be fixed from the initial conditions. Clearly, as we can see, the energy ($ E_{ST}\sim \eta_{max} $) associated with these BPS states grows \emph{linearly} with the size of the interval $ [0,\eta_{max}] $.

In the following, we consider the generalization of the above result (\ref{e86}) in the presence of $ \epsilon \ll 1$ deformations (\ref{e34}) to ST solution. The corresponding $ \eta $ equation modifies to,
\begin{eqnarray}
\eta'' = -\frac{\kappa^2 \epsilon}{8}-\eta'^2 \frac{\epsilon}{8}+\frac{\eta }{\left(4 \eta ^2+1\right)^2}\left( 1+\frac{\eta\left(-16 \eta ^4-32 \eta ^2+1\right) \epsilon }{8 \left(4 \eta ^2+1\right)}\right)(\chi'^2 +l^2 \sin^2\chi)
\end{eqnarray}
which by means of (\ref{e81}) further simplifies as,
\begin{eqnarray}
\eta'' = \frac{(\kappa^2 - \eta'^2)}{\eta (1+4 \eta^2)}+\frac{\epsilon  \left(\left(16 \left(\eta ^4+\eta ^2\right)-1\right) \eta'^2-\left(16 \left(\eta ^2+2\right) \eta ^2+3\right) \kappa ^2\right)}{8 \left(4 \eta ^2+1\right)^2}-\frac{\kappa^2 \epsilon}{8}-\eta'^2 \frac{\epsilon}{8}.
\label{e88}
\end{eqnarray}

Integrating (\ref{e88}) once, one can express the corresponding solution schematically as,
\begin{eqnarray}
\eta'_{DST} = \eta'_{ST}+\epsilon \tilde{\eta}'
\end{eqnarray}
where, $ \eta'_{ST} = \eta_{ST}^{-1}\sqrt{e^{2 c_1} \left(4 \eta_{ST} ^2+1\right)+\eta_{ST} ^2 \kappa ^2}$. Finding $ \tilde{\eta} $ is a bit tricky which we do not discuss here. This finally leads to the energy associated with the deformed ST solution,
\begin{eqnarray}
E_{DST}=E_{ST}-\frac{4\sqrt{\lambda}\kappa \epsilon}{\pi}\int_0^{\eta_{max}}\frac{\tilde{\eta}'}{\eta'^2_{ST}}d\eta + \cdots
\end{eqnarray} 
\subsubsection{Maldacena-Nunez solution}
Like before, we explore these extended string states in the context of MN solution (\ref{e38}). We first note down the corresponding $ \eta $ equation which in this case turns out to be,
\begin{eqnarray}
\eta'' = \frac{2 \kappa ^2 N^2}{\eta } 
\label{e91}
\end{eqnarray}
where we restrict ourselves to the holographic limit in which $ N$ is set to be infinity.

Using (\ref{e91}), finally the energy associated with the extended string turns out to be,
\begin{eqnarray}
E_{MN} = \frac{2\sqrt{2}\sqrt{\lambda}}{\pi}\int_0^{\eta_{max}}\frac{d\eta}{\sqrt{c-\log\eta}}+\mathcal{O}(1/N)
\end{eqnarray}
where $ c $ is the constant of integration.
\subsubsection{Revisiting the single kink and Uluru profile}
\label{revisiting strings in single kink and uluru}
We now consider the brane set up corresponding to two specific $ \mathcal{N}=2 $ linear quivers those have been discussed previously. Like before, we place the string soliton on the equatorial plane ($ \chi = \frac{\pi}{2} $) and start with the simple example of GM geometry corresponding to a \emph{single kink} potential (\ref{E53}) near region $ \sigma \sim 0 $. 

The corresponding $ \sigma $ equation turns out to be,
\begin{eqnarray}
\eta'^2 - \sigma'^2 = \frac{16 \kappa^2}{f^2_2}-\eta^2 l^2.
\label{e 131}
\end{eqnarray}

Combining (\ref{e 131}) with the Virasoro constraint (\ref{e81}) we get
\begin{eqnarray}
\eta' = \sqrt{\frac{16 \kappa^2}{f^2_2}-\eta^2 l^2}.
\label{e132}
\end{eqnarray}

Using (\ref{e132}), the energy associated with the string turns out to be,
\begin{eqnarray}
E_k &=&\frac{4\sqrt{\lambda}}{\pi}\left(P+1+\frac{l^2}{32\kappa^2}\int_0^{P+1}\eta^2 f^2_2 d\eta +\cdots \right)\nonumber\\ 
&= &\frac{4\sqrt{\lambda}}{\pi}(P+1)\left(1+\frac{l^2 \sigma^2_c}{16\kappa^2}(P+1)\left( (P+1)\partial_{\eta}\Lambda_{k}-2\int_0^{P+1}\frac{\eta^2(\eta^2 +3P^2)}{(P^2-\eta^2)^3}\right)  \right)
\label{e133}
\end{eqnarray}
which clearly exhibits a leading order correction to the string spectrum that is associated with the winding number ($ l $) of the string along the isometry direction of the two sphere. Interestingly enough, we see that like in the previous example of rotating strings, we encounter a pole precisely at the location ($ \eta =P $) of the flavor D6 branes. However, in case of rotating strings this pole was encountered in the expression for the R- charge. 

Considering the finite piece only, we finally note down the energy associated with the extended string configuration to be,
\begin{eqnarray}
E_k = \frac{4\sqrt{\lambda}}{\pi}(P+1)\left( 1+\frac{l^2 \sigma^2_c}{8 \kappa^2}(P+1)^4\left(c(k)-\frac{1}{(2P+1)^2} \right)\right).
\end{eqnarray} 

The second example we consider is that of the Uluru profile as discussed in (\ref{EE98}). A similar calculation in this particular case yields,
\begin{eqnarray}
E_u =\frac{4\sqrt{\lambda}}{\pi}\left( (2P+K)+\frac{l^2 \sigma^2_c}{8\kappa^2}\left( (2P+K)^3 c(u)+\int_0^{2P+K}\frac{\eta^2(\eta^2 +3P^2)}{(P^2-\eta^2)^3}\right) \right) 
\end{eqnarray}
exhibits similar pole(s) at the location ($ \eta \sim P $) of the D6 branes. 

Like before, one can further regulate these poles which results in a finite answer,
\begin{eqnarray}
E_u =\frac{4\sqrt{\lambda}}{\pi}(2P+K)\left( 1+\frac{l^2 \sigma^2_c}{8\kappa^2}(2P+K)^2\left( c(u)+\frac{1}{(K+P)^2(K+3P)^2}\right)  \right). 
\end{eqnarray}

Following our discussion in the previous Section, it is now straightforward to argue that the spacetime singularity near $ \eta \sim P $ fobids strings to actually reach the flavor D6 branes. In other words, strings require infinite energy to reach flavor D6 branes.

This can also be seen form a closer inspection of the sigma model Lagrangian (\ref{EE119}) 
\begin{eqnarray}
\mathcal{L}_P=f_2 (\sigma'^2 + \eta'^2)+V_{eff}
\end{eqnarray}
near the singularity which reveals a pole structure in the effective potential as we approach flavor D6 branes near $ \eta \sim P $,
\begin{eqnarray}
V_{eff}(\sigma \sim 0, P)= \frac{2 \sigma_c P^2 (P+1)^{1/2} l^2}{|\epsilon |^{3/2}}.
\end{eqnarray}
Notice that, the above potential has a slope which is different from the one for rotating strings (\ref{eee117}). This potential acts as a barrier that forbids folded strings to actually pass through flavor branes located at $ \eta =P $. Therefore, the resulting configuration could be thought of as a mixed ensemble of folded strings (floating around flavor D6 branes around $ \eta \sim P $) together with open strings that live on D6 branes.
\section{6d $\mathcal{N}=(1,0)$ SCFTs and type IIA dual}
The purpose of this part of analysis is to explore some hidden sectors of 6d $\mathcal{N}=(1,0)$ SCFTs by constructing dual semiclassical string states in type IIA supergravity with an $ AdS_7 $  factor \cite{Apruzzi:2013yva}, \cite{Cremonesi:2015bld}. We first focus on simply extended strings along the internal manifold ($ \mathcal{M}_3 $) and construct the corresponding states those carry zero angular momentum. 

In the second part of our analysis, we generalize our analysis by constructing string states with non zero angular momentum. These states correspond to rotation of the string along the isometry of $ S^2 $ in the internal manifold.

The Massive type IIA  background that we start with is considered to be dual to these 6d $\mathcal{N}=(1,0)$ SCFTs at strong coupling,
\begin{eqnarray}
ds^2&=&f_1(z)ds^2_{AdS_7}+f_2(z)dz^2+f_3(z)(d\chi^2 +\sin^2\chi d\xi^2)\\
B_2 &=& f_4(z)\sin\chi d\chi \wedge d\xi ~;~F_2=f_5(z)\sin\chi d\chi \wedge d\xi .
\end{eqnarray}

The functions $ f_i(z) $ are expressed in terms of another function $ \alpha(z) $ and its derivatives,
\begin{eqnarray}
f_1&=&8\sqrt{2}\pi \sqrt{-\frac{\alpha}{\alpha''}}~;~f_2 = \sqrt{2}\pi\sqrt{-\frac{\alpha''}{\alpha}}~;~f_3=f_2 \left( \frac{\alpha^2}{\alpha'^2 -2\alpha \alpha''}\right)\\
f_4&=&\pi \left( -z+\frac{\alpha \alpha'}{\alpha'^2 -2 \alpha \alpha''}\right) ~;~f_5(z)=\left(\frac{\alpha''}{162 \pi^2}+\frac{\pi F_0 \alpha \alpha'}{\alpha'^2 -2 \alpha  \alpha''} \right) 
\end{eqnarray}
where the function $ \alpha(z) $ corresponds to a class of supersymmetric solutions in Massive type IIA whose generic form is given by \cite{Nunez:2018ags},
\begin{eqnarray}
\alpha (z)=c_0 +c_1 z +c_2 z^2 -2\pi^3 F_0 z^3.
\end{eqnarray}
Here, the coefficients $ c_i $s are such that the function $ \alpha(z) $ is piece-wise continuous and differentiable that satisfies the boundary conditions namely, $ \alpha (0)=0 $ and $ \alpha(P+1)=0 $ for an interval $ 0\leq z \leq P+1 $.  Below we choose to work with different choices of these coefficients which correspond to different $ \mathcal{N}=(1,0) $ quivers in 6D.
\subsection{Long folded strings}
We first focus on extended folded string states with zero angular momentum. These strings have two segments in it which are stretched between $ z=0 $ and $ z=z_{max} $ and pass through flavor D8 branes those are localized at different positions along the $ z $ axis. 

Below we present a general algorithm that describes these sigma models.
We consider the string soliton to be sitting at the center of the global $ AdS_7 $ and is extended along the internal manifold ($ \mathcal{M}_3 $). In particular, we consider that the string wraps the isometry direction ($ \xi $) of $ S^2 \subset \mathcal{M}_3 $ while simultaneously stretches along the $ z $ axis. 

The ansatz we therefore propose is of the following form,
\begin{eqnarray}
t=\kappa \tau ~;~z=z(\tilde{\sigma})~;~ \chi =\frac{\pi}{2}~;~\xi = l \tilde{\sigma}.
\end{eqnarray}

The corresponding Lagrangian density turns out to be,
\begin{eqnarray}
\mathcal{L}_{P}=\kappa^2 f_{1}+f_2 z'^2 +l^2 f_3
\end{eqnarray}
which is supplemented by the Virasoro constraint(s) of the following form,
\begin{eqnarray}
T_{\tau \tau}&=&T_{\tilde{\sigma}\tilde{\sigma}}\nonumber\\
& = &-\kappa^2 f_{1}+f_2 z'^2 +l^2 f_3 = 0.
\end{eqnarray}

Taking the above as an input, the energy associated with these folded string configuration turns out to be,
\begin{eqnarray}
E_S = 8\sqrt{2}\sqrt{\lambda}\int_0^{z_{max}}\sqrt{\frac{\alpha'^2 - 2\alpha \alpha''  }{8\alpha'^2 -15 \alpha \alpha''}}dz.
\label{e145}
\end{eqnarray}
where for simplicity we set, $ \kappa =1=l $.

Below, we enumerate the above formula (\ref{e145}) taking various examples of $ \mathcal{N}=(1,0) $ linear quivers in 6d.
\subsubsection{Quiver I}
The first example we consider is that of a long quiver without any \emph{plateau}.  The corresponding gauge group $ SU(N)\times SU(2N)\times\cdots \times SU(PN) $ is closed by placing a flavor group $ SU((P+1)N) $ at the end of the node. 

In the holographic limit, the corresponding type IIA supergravity solution is characterised by the function \cite{Nunez:2018ags},
\begin{equation}
 - \frac{\alpha(z)}{81\pi^2 N } =
    \begin{cases}
      a_1 z+\frac{z^3}{6} &  0\leq  z \leq 1\\
      (k a_1 +\frac{k^3}{6})+(a_1+\frac{k^2}{2})(z-k)+\frac{k}{2}(z-k)^2 +\frac{1}{6}(z-k)^3 & k\leq z \leq (k+1)\\
      (P a_1 +\frac{P^3}{6})+(a_1+\frac{P^2}{2})(z-P)+\frac{P}{2}(z-P)^2-\frac{P}{6}(z-P)^3 & P \leq z \leq P+1
    \end{cases} 
    \label{e146}      
\end{equation}
where $ a_1=-\frac{1}{6}(P^2+2P) $ and $ k=1,..,P-1 $ where $ P\gg 1 $.

Using (\ref{e146}), one can estimate the integral (\ref{e145}) piece-wise which finally yields (with string $ \alpha'_S $ coupling equal to one),
\begin{eqnarray}
E_S = \frac{9P}{2 } +\mathcal{O}(1/P). 
\label{e153}
\end{eqnarray} 

Notice that, in the holographic limit ($ P \gg 1 $), the energy of the string is proportional to the maximum rank of the $ SU(N) $ color group. However, considering the fact that the folded string has two segments in it, the energy corresponding to each segment roughly equals to $ \Delta_{open}= 2P $ which is reminiscent of the baryon operators dual to D0 strings \cite{Bergman:2020bvi}.

For the plateau-less case, as we have considered above, the flavor D8 branes are located nearly at the end point ($ z=P $) of the long string. The folded string that we study here just passes through the stack of D8 branes at $ z=P $ and then again ends on them.

Below we compute Page charges associated with the brane set up.

We start by looking at the Page charge associated with the NS5 brane configuration,
\begin{eqnarray}
Q_{NS5}=-\frac{1}{4\pi^2}\int_{\mathcal{M}_3} H_3
\end{eqnarray}
where $ H_3 =dB_2 $. A straightforward computation reveals,
\begin{eqnarray}
Q_{NS5}=-\frac{1}{\pi}(f_4 (P+1)-f_4(0))= P +\mathcal{O}(1/P).
\end{eqnarray}

On the other hand, a straightforward computation of the Page charge corresponding to D6 branes reveals \cite{Nunez:2018ags},
\begin{eqnarray}
Q_{D6}=-\frac{1}{2\pi}\int_{S^2} F_2-F_0 B_2.
\end{eqnarray}

It turns out that the major contribution to the D6 brane charge comes from the plateauless region between $ 1\leq z \leq P-1 $. Adding all these contributions together and taking the limit $ P \gg 1 $ we find the total D6 brane charge,
\begin{eqnarray}
Q_{D6}^{(Total)}= \frac{N P^2}{2}+\mathcal{O}(1/P).
\end{eqnarray}

Combining these pieces together we find,
\begin{eqnarray}
E_S = \frac{9}{2}Q_{NS5} = \sqrt{\frac{2}{N}} \sqrt{Q^{(Total)}_{D6}}.
\end{eqnarray}
\subsubsection{Quiver II}
The second example that we focus here the folded string actually passes through the stack of flavor D8 branes at locations $ z=q $ and  $ z=N-q $. This is an example of a quiver where there is no plateau region for $ z\leq q $ (which therefore corresponds to the \emph{massive} case with $ F_0 \neq 0 $), then there is a plateau region (with $ F_0 =0 $) for $ q\leq z \leq (N-q) $ and finally the rank of the gauge group starts falling for $ z\geq N-q $.

The corresponding supergravity background is characterized by the function \cite{Bergman:2020bvi}
\begin{equation}
 - \frac{\alpha(z)}{9\pi^2 /6 } =
    \begin{cases}
       z(3k(z-N)+n(3q(q-N)+z^2)) &  0\leq  z \leq q\\
      n q^3-3N(k+nq)z+3(k+nq)z^2 & q\leq z \leq N-q\\
      (N-z)(-3kz+n(3q(q-N)+(N-z)^2)) &  z \geq N-q
    \end{cases} 
    \label{e148}      
\end{equation}
where, $ k $ and $ n $ are the ranks of the corresponding flavor groups.

In order to evaluate the corresponding integral (\ref{e145}), we take into account the large $ N $ limit such that the ratios $ \frac{q}{N}$ and $ \frac{k}{N} $ are fixed. This finally yields (with string $ \alpha'_S $ coupling equal to one),
\begin{eqnarray}
E_S= \frac{395}{96} N+\mathcal{O}(q/N).
\end{eqnarray}
which shows that in the large $ N $ limit the operator dimension ($ \Delta = E_S $) in the dual $ \mathcal{N}=(1,0) $ SCFT grows linearly with $ N $. 

Notice that, each segment of the folded string could be thought of as an open string whose end points are eventually glued together. The energy associated with each of these segments therefore turns out to be $ \Delta_{open} = 2 N  $. Therefore, in the large $ N $ limit, the dominant contribution to the spectrum arises due to these open string segments those are extended between the flavor D8 branes located at $ z=q $ and $ z=N-q $.

Finally, we note down the Page charges associated with these brane set up that we are discussing here. We first compute the Page charge associated with NS5 branes which in the large $ N $ limit turns out to be,
\begin{eqnarray}
Q_{NS5}= N +\mathcal{O}(1/N).
\end{eqnarray}

On the other hand, the Page charge corresponding to D6 branes becomes
\begin{eqnarray}
Q_{D6}= \begin{cases}
       6k &  0\leq  z \leq q\\
      6 (k+n q)& q\leq z \leq N-q\\
      6 (k+n N)& z \geq N-q.
    \end{cases} 
\end{eqnarray}

Adding all these pieces together, the total D6 brane charge may be obtained as,
\begin{eqnarray}
Q^{(Total)}_{D6}= 6n N +\mathcal{O}(1/N).
\end{eqnarray}

These together suggest that the dispersion relation corresponding to these extended semi-classical strings may be put in the form,
\begin{eqnarray}
E_S = \frac{395}{96}Q_{NS5}= \frac{395}{576 n}Q^{(Total)}_{D6}. 
\end{eqnarray}
\subsection{Rotating string solutions}
We now construct the semiclassical string states with nonzero angular momentum where the string is considered to be rotating along the isometry of $ S^2 \subset \mathcal{M}_3 $.

To start with, we choose an ansatz of the following form,
\begin{eqnarray}
t=\kappa \tau ~;~z=z(\tilde{\sigma})~;~ \chi =\frac{\pi}{2}~;~\xi = \omega \tau.
\end{eqnarray}

The corresponding sigma model Lagrangian turns out to be,
\begin{eqnarray}
\mathcal{L}_P =\kappa^2 f_1 -\omega^2 f_3 +f_2 z'^2
\end{eqnarray}
which is supplemented by the Virasoro constraints of the following form,
\begin{eqnarray}
T_{\tau \tau}&=&T_{\tilde{\sigma}\tilde{\sigma}}\nonumber\\
& = &-\kappa^2 f_{1}+f_2 z'^2 +\omega^2 f_3 = 0.
\end{eqnarray}

Finally, we note down the energy and the angular momentum associated with the folded string configuration which turns out to be (setting string $ \alpha'_S $ coupling to unity),
\begin{eqnarray}
\label{e168}
E_S &=&8\sqrt{2}\int_0^{z_{max}}\sqrt{\frac{\alpha'^2 - 2\alpha \alpha''  }{8\alpha'^2 -15 \alpha \alpha''}}dz\\
J &=&-\sqrt{2} \int_0^{z_{max}}\frac{\alpha \alpha''}{\sqrt{8 \alpha'^4 -31 \alpha \alpha'^2 \alpha'' + 30 \alpha^2 \alpha''^2}}
\label{e169}
\end{eqnarray}
where we set, $ \kappa =\omega =1 $ without any further loss of generality.

Below, we estimate the above charges (\ref{e168}), (\ref{e169}) corresponding to two different quivers (\ref{e146}) and (\ref{e148}) as mentioned above.
\subsubsection{Quiver I}
The energy corresponding to quiver I (\ref{e146}) remains the same as before
\begin{eqnarray}
E_S = \frac{9P}{2 } +\mathcal{O}(1/P).
\label{e170}
\end{eqnarray}

On the other hand, the angular momentum takes the following form,
\begin{eqnarray}
J = P + \mathcal{O}(1/P). 
\label{e171}
\end{eqnarray}

The above two equations (\ref{e170}) and (\ref{e171}) may be combined together to yield,
\begin{eqnarray}
\label{eee174}
E_S -J  = \frac{7}{2}Q_{NS5}= \frac{7}{\sqrt{2N}}\sqrt{Q_{D6}^{(Total)}}.
\end{eqnarray}
\subsubsection{Quiver II}
We now compute the string energy as well as the angular momentum corresponding to quiver II (\ref{e148}). The computation is eventually quite straightforward and similar to the previous computations which finally yields,
\begin{eqnarray}
E_S&= & \frac{395}{96} N+\mathcal{O}(q/N)\\
J & = &\frac{N}{2}+\mathcal{O}(q/N).
\end{eqnarray}

The above two could be combined together to yield,
\begin{eqnarray}
\label{eee177}
E_S -J = \frac{347}{96}Q_{NS5}= \frac{347}{576n}Q^{(Total)}_{D6}.
\end{eqnarray}

Combining our results for quiver I and quiver II, the general form of the dispersion relation turns out to be,
\begin{eqnarray}
\label{eee178}
E_S -J = g Q^{\frac{1}{p}}_{D6}
\end{eqnarray}
where, $ g $ is the constant of proportionality that depends on the choice of the quiver in Fig.\ref{uluru}. On the other hand, the integer number $ p $ is quite unique to the choice of the quiver. For example, it takes the value $ 2 $ for the single kink profile Fig.\ref{uluru}a whereas its value turns out to be $ 1 $ corresponding to quiver II as depicted in Fig.\ref{uluru}b.
\subsection{Remarks on effective potential}
Unlike the previous example of $ \mathcal{N}=2 $ linear quivers, the operator spectrum in the dual $ \mathcal{N}=(1,0) $ SCFT does not exhibit any pole structure as the string approaches the flavor D8 branes. Below, we argue that this finite spectrum is due to the \emph{regular} behaviour of the effective string potential near the flavor D8 branes. 

The sigma model Lagrangian has the generic structure of the form,
\begin{eqnarray}
\mathcal{L}_P =f_2 z'^2 + V_{eff}(z)
\end{eqnarray}
where the effective potential could be formally expressed as,
\begin{eqnarray}
V_{eff}(z) = \sqrt{2}\pi \sqrt{-\frac{\alpha}{\alpha''}}\left(8 \gamma_1 + \gamma_2 \frac{\alpha'' \alpha}{\alpha'^2 -2\alpha \alpha''} \right) 
\label{e177}
\end{eqnarray}
with $ \gamma_{1,2} $ as some constant parameters of the theory.

Considering quiver I (\ref{e146}), it is now straightforward to obtain
\begin{eqnarray}
\frac{V_{eff} (z)}{\frac{2 \pi}{\sqrt{3}}}=  4  \gamma _1 P^2+ \left(8 \gamma _1-\gamma _2\right) P+\left(\gamma _2 \left(-3 z^2+z-2\right)-2 \gamma _1 (z+1)^2\right)+\mathcal{O}(1/P).
\end{eqnarray}

This clearly shows that the effective potential is regular at $ z = P $ 
\begin{eqnarray}
\frac{V_{eff} (z \sim P)}{\frac{2 \pi}{\sqrt{3}}}= \left(2 \gamma _1-3 \gamma _2\right) P^2+4  \gamma _1 P-2 \left(\gamma _1+\gamma _2\right).
\end{eqnarray}
Therefore the string passes smoothly across the flavour D8 branes (located at $ z=P $) yielding a finite spectrum at strong coupling.

A similar argument holds for quiver II (\ref{e148}) which we therefore do not repeat here. 
\subsection{Coupling to the $ B $ field}
The string soliton we have considered so far does not couple to the NS-NS two form ($ B_2 $) for obvious reasons. However, as we shall see, that the dynamics becomes almost unsolvable once the string is coupled to the $ B $ field. 

In order to incorporate the $ B $ field into the picture, we propose an ansatz of the following form,
\begin{eqnarray}
t=\kappa \tau ~;~z=z(\tilde{\sigma})~;~ \chi =\chi (\tilde{\sigma})~;~\xi = \omega \tau.
\end{eqnarray}

This yields a Lagrangian of the following form,
\begin{eqnarray}
\mathcal{L}_P = \kappa^2 f_1 -\omega^2 f_3 \sin^2\chi +f_2 z'^2 +f_3 \chi'^2 -2\omega f_4 \sin\chi \chi'.
\end{eqnarray}

The resulting equations of motion are given by
\begin{eqnarray}
2f_2 z'' &=&\kappa^2 \partial_z f_1 -\omega^2 \partial_z f_3 \sin^2\chi -\partial_z f_2 z'^2+\partial_z f_3 \chi'^2 -2\omega \partial_z f_4 \sin\chi \chi'\\
f_3 \chi'' &=&-\partial_z f_3 z' \chi' +\omega \partial_z f_4 z' \sin\chi -\omega^2 f_3 \sin\chi \cos\chi.
\label{ee183}
\end{eqnarray}

Using (\ref{ee183}) as well as the Virasoro constraint this finally leads to
\begin{eqnarray}
z'(\tilde{\sigma}) =\sqrt{-\frac{8  \alpha}{\alpha''}}\left( 1+\frac{1}{8\sqrt{2}\pi}\sqrt{-\frac{\alpha''}{\alpha}}\mathcal{K}\right) ^{1/2} 
\end{eqnarray}
where we set, $ \kappa =\omega =1 $ together with the fact that,
\begin{eqnarray}
\mathcal{K}=\int_0^{2\pi}(\sin\chi (2\partial_z f_4 z' \chi' + \partial_z f_3 z'\sin\chi )-\partial_z f_3 z'\chi'^2) d\tilde{\sigma}.
\end{eqnarray}

This finally leads to the energy (or mass) associated with the long string as,
\begin{eqnarray}
\label{ee186}
E_S =8\pi \int_{0}^{z_{max}}\frac{dz}{\sqrt{1+\frac{1}{8\sqrt{2}\pi}\sqrt{-\frac{\alpha''}{\alpha}}\mathcal{K}}}.
\end{eqnarray} 
Clearly, the evaluation of (\ref{ee186}) is a bit tricky due the presence of the nontrivial coupling between $ \chi $ and $ z $ fields. 

Notice that, the above integral (\ref{ee186}) simplifies enormously once we decouple the string from background NS fluxes. One way to do it is to set $ \chi = \chi_0  \sim $ constant which sets, $ \mathcal{K}\sim f_3(z_{max})-f_3(0) \sim 0 $. As a result, the energy of the string effectively becomes proportional to the size of the quiver- as observed earlier. 

Following similar steps, we obtain an expression for the angular momentum
\begin{eqnarray}
\label{eqj}
J = \pi \int_{0}^{z_{max}}\left( \frac{\alpha'' \alpha \sin^2\chi}{\alpha'^2 -2\alpha \alpha''}\right) \frac{dz}{\sqrt{1+\frac{1}{8\sqrt{2}\pi}\sqrt{-\frac{\alpha''}{\alpha}}\mathcal{K}}}.
\end{eqnarray}

Combining (\ref{ee186}) and (\ref{eqj}), we find a dispersion relation
\begin{eqnarray}
E_S - J =\pi \int_{0}^{z_{max}}\left( \frac{8\alpha'^2 - 16 \alpha \alpha'' -\alpha'' \alpha \sin^2\chi}{\alpha'^2 -2\alpha \alpha''}\right) \frac{dz}{\sqrt{1+\frac{1}{8\sqrt{2}\pi}\sqrt{-\frac{\alpha''}{\alpha}}\mathcal{K}}},
\end{eqnarray}
whose analytic evaluation is indeed a quite non-trivial task for the same reasons as mentioned above.
\section{2d $\mathcal{N}=(0,4)$ SCFTs and type IIA dual}
To conclude our analysis, as a last example, we explore the dynamics of extended folded string configurations over a new class of massive type IIA $ AdS_3 $ supergravity backgrounds those are dual to $ \mathcal{N}=(0,4) $ SCFTs in $ 2d $ \cite{Lozano:2019emq}-\cite{Lozano:2019ywa}.

The NS-NS sector of the full type IIA solution could be formally expressed as \cite{Lozano:2019zvg},
\begin{eqnarray}
ds^2 &=& f_1 (\rho)ds^2_{AdS_3}+f_2(\rho)ds^2_{S^2}+f_3(\rho)d\rho^2 + f_4(\rho)ds^2_{CY_2}\\
B_2 &=&\frac{1}{2}(2 k \pi - \rho +\frac{u \partial_{\rho}u}{4 h_4 h_8 +(\partial_{\rho}u)^2})\sin \chi d\chi \wedge d\xi \\
e^{-\Phi}&=&\frac{h^{\frac{3}{4}}_{8}}{2h^{\frac{1}{4}}_4 \sqrt{u}}\sqrt{4 h_4 h_8 +(\partial_{\rho}u)^2}
\end{eqnarray}
where the individual metric functions ($ f_i$) could be schematically expressed as,
\begin{eqnarray}
f_1(\rho)&=&\frac{u}{\sqrt{h_4 h_8}}~~;~~f_2(\rho)=f_1 \frac{h_4 h_8}{4 h_4 h_8 +(\partial_{\rho}u)^2}\\
f_3(\rho)&=&\frac{\sqrt{h_4 h_8}}{u}~~;~~f_4 (\rho)=\sqrt{\frac{h_4}{h_8}}.
\end{eqnarray}

Following Bianchi identity as well as background SUSY, the functions are constrained to be linear as well as  piecewise continuous,
\begin{eqnarray}
\label{eq201}
\partial_{\rho}^2h_4(\rho)=0~~;~~\partial_{\rho}^2h_8(\rho)=0~~;~~\partial_{\rho}^2u=0
\end{eqnarray}
which thereby results in a infinite family of solutions corresponding to different $ \mathcal{N}(0,4) $ linear quivers at strong coupling. 

Here $ \rho $ (is the coordinate associated to the internal space) stands for the field theory direction which increases through an interval of $ 2\pi $. The quiver begins at $ \rho =0 $ and ends at $ \rho =2\pi (P+1) $ where $ P\gg1 $ in the supergravity approximation.

For the purpose of our present analysis, we will focus on the stringy dynamics associates to $ S^2 $ of the internal manifold while we keep the directions associated with $ CY_2 $ to be fixed. The corresponding Hanany-Witten (HW) configuration that we have in mind is that of NS-D2-D4-D6-D8 brane set up where the NS five branes are localized along the $ \rho $ coordinate of the internal manifold while the color D6 branes are extended (between different NS5 branes) along the $ \rho $ coordinate. On the other hand, we have flavor D8 branes localized along $ \rho $ and thereby changing the slope of the quiver in the corresponding locations.

The Page charges those are relevant for the present solitonic configuration are the fluxes (through transverse directions associated to D6 and D8 brane configurations) of the following form \cite{Lozano:2019zvg},
\begin{eqnarray}
Q_{D8}&=&2\pi \partial_{\rho}h_8\\
Q_{D6}&=&\frac{1}{2\pi}(h_8 -(\rho - 2\pi k)\partial_{\rho}h_8)
\end{eqnarray}
which is defined for an interval $ 2\pi k \leq \rho \leq 2\pi (k+1) $ where $ k=1,\cdots,P-1 $.
\subsection{Long strings without rotation}
Like before, we start by considering the so called long string states without rotation. These strings are supposed to be extended through different flavor D8 branes localized along the $ \rho $ coordinate of the internal manifold.

To study these long folded strings, we propose an ansatz of the following form,
\begin{eqnarray}
t = \tau ~~;~~\rho = \rho (\tilde{\sigma})~~;~~\chi =\frac{\pi}{2}~~;~~\xi =\ell \tilde{\sigma}
\end{eqnarray}
that essentially decouples the string from the background NS fluxes and thereby simplify the analysis enormously.

The corresponding Lagrangian density is given by
\begin{eqnarray}
\mathcal{L}_P =f_1 + \ell^2 f_2 +f_3 \rho'^2
\end{eqnarray}
supplemented by the Virasoro constraints of the following form, 
\begin{eqnarray}
T_{\tau \tau}&=&T_{\tilde{\sigma}\tilde{\sigma}}\nonumber\\
& = &- f_{1}+f_3 \rho'^2 +\ell^2 f_2 = 0.
\end{eqnarray}

Finally, we note down the energy (or mass) of the long string to be,
\begin{eqnarray}
E_S = \frac{1}{\pi}\int_0^{\rho_{max}}\sqrt{\frac{4h_4h_8 +(\partial_{\rho}u)^2}{3h_4h_8 +(\partial_{\rho}u)^2}}d\rho
\label{e198}
\end{eqnarray}
where we set the winding number $ \ell=1 $.

Below we estimate (\ref{e198}) for a family of type IIA solutions \cite{Lozano:2019zvg} those correspond to different $ \mathcal{N}=(0,4) $ linear quivers in $ 2d $.
\subsubsection{Quiver I}
The first example of the quiver that we consider is characterized by the function $ u=\frac{b_0}{2\pi}\rho $ together with the following type IIA solutions
\begin{eqnarray}
\label{e199}
h_8 (\rho)=\begin{cases}
       \frac{\nu}{2\pi}\rho &  0 \leq \rho \leq 2\pi\\
      \nu & 2\pi \leq \rho \leq 2\pi P\\
      \frac{\nu}{2 \pi}(2\pi (P+1)-\rho) & 2\pi P \leq \rho \leq 2\pi (P+1),
    \end{cases} 
\end{eqnarray}
\begin{eqnarray}
\label{e200}
h_4 (\rho)= \begin{cases}
       \frac{\beta}{2\pi}\rho &  0 \leq \rho \leq 2\pi\\
      \beta & 2\pi \leq \rho \leq 2\pi P\\
      \frac{\beta}{2 \pi}(2\pi (P+1)-\rho) & 2\pi P \leq \rho \leq 2\pi (P+1)
    \end{cases} 
\end{eqnarray}
where the holographic limit corresponds to setting the ranks of the $ SU(N) $ color groups to be $ \nu $, $ \beta \rightarrow \infty $ together with $ P \gg 1 $ \cite{Lozano:2019zvg}.

Using (\ref{e199}) and (\ref{e200}) we estimate the integral (\ref{e198}) piecewise which finally yields\footnote{With string $ \alpha'_S $ coupling equal to one.},
\begin{eqnarray}
E_S = \frac{4   P}{\sqrt{3}}.
\label{e201}
\end{eqnarray} 

On the other hand, a direct computation on the Page charges reveals,
\begin{eqnarray}
Q_{D8}& = &\nu \\
Q_{D6}&= & 2 \nu P^2.
\label{e203}
\end{eqnarray}

Using (\ref{e203}), one can further rearrange the expression for the energy/mass of the long folded string which comes out to be,
\begin{eqnarray}
E_S = \frac{2}{\sqrt{6 \nu} }Q^{1/2}_{D6}.
\end{eqnarray} 
\subsubsection{Quiver II}
The second example of the quiver that we consider is characterized by the following functions 
\begin{eqnarray}
\label{e205}
h_8 (\rho)= \begin{cases}
       \frac{\nu}{2\pi}\rho &  0 \leq \rho \leq 2\pi P\\
      \frac{\nu P}{2 \pi}(2\pi (P+1)-\rho) & 2\pi P \leq \rho \leq 2\pi (P+1),
    \end{cases} 
\end{eqnarray}
\begin{eqnarray}
\label{e206}
h_4 (\rho)= \begin{cases}
       \frac{\beta}{2\pi}\rho &  0 \leq \rho \leq 2\pi P\\
      \frac{\beta P}{2 \pi}(2\pi (P+1)-\rho) & 2\pi P \leq \rho \leq 2\pi (P+1)
    \end{cases} 
\end{eqnarray}
which corresponds to a quiver with linearly increasing rank associated with the color group. These quivers are closed by placing flavour D4/D8 branes after P nodes.

The corresponding Page charges are given by,
\begin{eqnarray}
\label{e207}
Q_{D8}& = & \nu P \\
Q_{D6}&= & \nu P^3.
\label{e208}
\end{eqnarray}

In terms of (\ref{e207})-(\ref{e208}), one can further re-express the energy/mass of the string as,
\begin{eqnarray}
E_S = \frac{4  P}{\sqrt{3}}= \frac{4}{\sqrt{3}\nu^{1/3}}Q^{1/3}_{D6}.
\end{eqnarray}
\subsubsection{Quiver III}
The third example we consider is that of quiver corresponding to (Fig.\ref{uluru}b). The associated (massive) type IIA geometry is characterized by the functions,
\begin{eqnarray}
\label{e205}
h_8 (\rho)= \begin{cases}
       \frac{\nu}{2\pi}\rho &  0 \leq \rho \leq 2\pi K\\
       \nu K & 2\pi K \leq \rho \leq 2\pi (K +q)\\
      \frac{\nu K}{2 \pi (P+1-K - q)}(2\pi (P+1)-\rho) & 2\pi (K+q) \leq \rho \leq 2\pi (P+1),
    \end{cases} 
\end{eqnarray}
\begin{eqnarray}
\label{e206}
h_4 (\rho)= \begin{cases}
       \frac{\beta}{2\pi}\rho &  0 \leq \rho \leq 2\pi K\\
       \beta K &2 \pi K \leq \rho \leq 2\pi (K +q)\\
      \frac{\beta K}{2 \pi(P+1-K -q)}(2\pi (P+1)-\rho) & 2\pi (K + q) \leq \rho \leq 2\pi (P+1).
    \end{cases} 
\end{eqnarray}

The corresponding Page charges are given by,
\begin{eqnarray}
\label{e212}
Q_{D8}& = & \nu K\\
Q_{D6}& = &    K P^2 \nu 
\label{e213}
\end{eqnarray}
in the limit in which $ P \gg q \gg K \gg 1 $.

Finally, using (\ref{e212})-(\ref{e213}) the energy of the long string turns out to be,
\begin{eqnarray}
E_S = \frac{4}{\sqrt{3 K \nu}}Q^{1/2}_{D6}.
\end{eqnarray} 
\subsection{Rotating string solutions}
We now generalise the above string solutions in the presence of non-zero rotation along the isometry of $ S^2 $ of the internal manifold. To this end, we choose to work with the string embedding of the following form,
\begin{eqnarray}
 t=\tau ~~;~~\rho = \rho (\tilde{\sigma})~~;~~\chi =\frac{\pi}{2}~~;~~\xi =\omega \tau
\end{eqnarray}
which results in the sigma model Lagrangian of the following form,
\begin{eqnarray}
\mathcal{L}_{P}=f_1 -\omega^2 f_2 +f_3 \rho'^2
\end{eqnarray}
together with the Virasoro constraint(s),
\begin{eqnarray}
T_{\tau \tau}&=&T_{\tilde{\sigma}\tilde{\sigma}}\nonumber\\
& = &- f_{1}+f_3 \rho'^2 +\omega^2 f_2 = 0.
\end{eqnarray}

The corresponding angular momentum associated with the long strings may be expressed as,
\begin{eqnarray}
J = \frac{1}{\pi}\int_0^{\rho_{max}}\frac{h_4 h_8}{\sqrt{(3h_4h_8 +(\partial_{\rho}u)^2)(4h_4h_8 +(\partial_{\rho}u)^2)}}d\rho
\label{e218}
\end{eqnarray}
where we set, $ \omega =1 $ without any loss of generality.

Evaluating (\ref{e218}) piecewise for each of the above quivers we find,
\begin{eqnarray}
\label{e219}
E_S -J = \begin{cases}
        \sqrt{\frac{3}{2 \nu}}Q^{1/2}_{D6}&  \text{Quiver I}\\
        \sqrt{3}\frac{Q_{D6}^{1/3}}{\nu^{1/3}} & \text{Quiver II}\\
      \frac{\sqrt{3}}{\sqrt{K \nu}}Q^{1/2}_{D6}& \text{Quiver III}.
    \end{cases} 
\end{eqnarray}
\subsection{Some further remarks}
Some interesting observations are the following. Both for for $ 6d $ $ \mathcal{N}=(1,0) $ SCFTs and $ 4d $ $ \mathcal{N}=(0,4) $ SCFTs, we find evidence that long operators are stable and therefore do not fragment into shorter operators. The corresponding long string states in the dual type IIA supergravity pass through the flavour branes without being breaking apart into smaller segments. This we identify as a generic feature of massive type IIA backgrounds.

Like in the $ 6d $ case, the long strings dual to operators corresponding to $ 4d $ $ \mathcal{N}=(0,4) $ linear quivers experience a potential that is regular near the location of flavour branes,
\begin{eqnarray}
\label{e220}
V_{eff} = \begin{cases}
      \frac{1}{2} \pi  a \sqrt{\frac{1}{\beta \nu}} (\gamma +4) P &  \text{Quiver I}\\
     \pi^2  a \sqrt{\frac{1}{\beta \nu}} (\gamma +4)   & \text{Quiver II}\\
   = \sqrt{\frac{1}{\beta \nu}}~,~ \frac{a \pi(4 +\gamma)(K+q)}{2K} \sqrt{\frac{1}{\beta \nu}} & \text{Quiver III}
    \end{cases} 
\end{eqnarray}
where, we have considered the function $ u(\rho) =a \rho $ with $ a $ being a (constant) proportionality factor together with, $ \gamma =\pm 1 $. As a general remark, one can therefore conclude that in the strict holographic limit ($ \nu,\beta \rightarrow \infty $), the string does not ``feel" any potential/ the presence of flavour branes as it passes through the stack of flavour branes. 

Like before, we also observe a general structure for the dispersion relation which may be put in the form,
\begin{eqnarray}
\label{eee224}
E_S -J = f Q_{D6}^{\frac{1}{p+1}}
\end{eqnarray}
where $ p $ is an integer that takes the value $ 2 $ for single kink (Fig.\ref{uluru}a) and $ 1 $ for linear quiver as depicted in Fig\ref{uluru}b. Therefore, comparing with our previous results in the $ 6d $ case, we notice that the value of $ p (=1,2) $ essentially characterises the quiver that we are dealing with. In other words, its value is unique to the choice of quivers we are working with.

\section{Summary and final remarks}
We now conclude our paper by summarizing all the key findings together with an interpretation from the perspective of the dual SCFTs. 

The present paper explores the possibilities of existence of long/heavy operators in a class of stongly coupled SCFTs those are living in diverse dimensions and accompanied with different amount of supersymmetries. These are (i) $ \mathcal{N}=(1, 0) $ SCFTs in $ 6d $, (ii) $ \mathcal{N}=2 $ SCFTs in $ 4d $ and (iii) $ \mathcal{N}=(0,4) $ SCFTs in $ 2d $. 

The analysis is quite similar in spirit to those performed in the context of $ 4d $ $ \mathcal{N}=4 $ SYM in the sense that it reinvestigates some of the major issues (in the context of these new classes SCFTs) those have been investigated previously in the context of $ \mathcal{N}=4 $ SYM in $ 4d $. One of the notable differences between these classes of SCFTs and $ 4d $ $ \mathcal{N}=4 $ SYM is that the later has a Lagrangian description while the former do not.

We address this question into two parts- first we show that it is indeed possible to construct such long operators (those behave like a \emph{closed} chain of fields/ d.o.f) in the SCFTs mentioned above and second it is also possible to quantify the spectrum associated with these class of operators at strong coupling.

Both the answers come from studying the dynamics associated with \emph{long} folded string configurations those probing type IIA supergravity backgrounds with topology $ AdS_p \times \mathcal{M}^q $ where $ \mathcal{M}^q  $ is some compact internal space where the stringy dynamics is mostly confined. These long strings extend through the flavour branes located along internal direction (which is also known as the holographic/field theory direction) associated to $ \mathcal{M}^q  $ and covers the entire space. 

From the perspective of the Hanany-Witten brane set up, these folded strings are therefore extended along (one of the world-volume direction of) the color branes and passes through flavour branes along the transverse direction (see Fig. \ref{folded-string}a). These folded strings are therefore dual to a class of (gauge invariant) \emph{heavy} operators/ long (closed) chain of adjoint matter (corresponding to the vector multiplet of the $ SU(N_c) $ color group) those belong to some particular sector of the $ SU(N_c) $ gauge group and contains some $ SU(N_f) $ flavour degrees of freedom as well. 

The appearance of these additional flavour d.o.f. is an artefact of the passing of the folded string through localised flavour branes along the internal manifold ($ \mathcal{M}^q  $). In other words, the insertion of these flavour fields (along the chain of adjoint matter) corresponds to the location of flavour branes in the $ 10 $d spacetime.

As a schematic illustration of the above argument, one could think of a dual operator of the following form,
\begin{eqnarray}
\label{e225}
\mathcal{O}^{f_m f_{m+n}}\sim ~~~~~~~~~~~~~~~~~~~~~~~~~~~~~~~
~~~~~~~~~~~~~~~~~~~~~~~~~~~~~~~~~~~~~~~~~~~~~~~~~~~~~~~~~~~~~~~
~~~~~
 \nonumber\\
tr (\Phi^{a_1 a_2}\Phi^{a_2 a_3}\cdots \Phi^{a_{k-1}a_k}\psi^{a_k f_m}\Phi^{a_{k+1}a_{k+2}}\cdots\Phi^{a_{k+n-1}a_{k+n}}\psi^{a_{k+n} f_{m+n}}\Phi^{a_{k+n+1}a_{k+n+2}}\cdots\Phi^{a_{k+k'}a_{1}})
\end{eqnarray}
where the trace is over the adjoint indices $\lbrace a_{i} \rbrace$ and $ f_{m} $s are the flavour indices corresponding to the location of these flavour branes at $ d_{m} $ and $ d_{m+n} $ (Fig.\ref{uluru}b).

Given the above picture - the present analysis reveals the following:\\
$ \bullet $ In the case of $ \mathcal{N}=2 $ linear quivers, such states are eventually \emph{fragmented} into smaller states containing only chain of adjoint matter plus some additional matter content living on the flavour branes. These additional matter content is an artefact of the fluctuations of open strings living on flavour D6 branes. For example, given the above example (\ref{e225}), one could think of a fragmentation of the form
\begin{eqnarray}
\mathcal{O}^{f_m f_{m+n}} \sim \mathcal{O}^{f_m }\mathcal{O}^{ f_{m+n}}
\end{eqnarray}
where $ f_{m} $ and $ f_{m+n} $ are the indices associated to open strings living on flavour D6 branes.

In the dual string theory picture this corresponds to the fact that the folded string can actually never connects two different stack of flavour D6 branes located at two distinct spacetime points along the holographic axis namely $ d_m $ and $ d_{m+n} $. In other words, they get fragmented into smaller strings across these flavour branes.

 An equivalent way of saying this is that the \emph{effective} string potential shows an infinite deep near the location of these flavour branes. As a result, one has a picture of closed strings floating around flavour branes (but never reaching them) plus some open string states living on flavour D6 branes (see Fig. \ref{folded-string}b).\\
$ \bullet $ In case of $ \mathcal{N}=(1,0) $ linear quivers in $ 6d $ and $ \mathcal{N}=(0,4) $ linear quivers in $ 2d $ such heavy operators (containing chain of both adjoint as well flavour d.o.f.) turn out to be stable. In other words, we do not see such fragmentation to occur at strong coupling. In the dual string theory picture this refers to the fact that long string states can pass through flavour branes without causing any additional effects which is an artefact of the smoothness of the effective string potential/or the geometry near the location of these flavour D8 branes.

To conclude, we must say that based on purely string theory calculations, the present paper actually conjectures the existence of above operators for a large class of SCFTs those can exist in diverse dimensions (starting from $ 6d $ down to $ 2d $) with different amount of SUSY. These are therefore results that we understand purely from the ground of string theory and should be reproducible from the perspective of the dual field theory. 

We hope to be able to report some of these results in the near future. \\

{\bf {Acknowledgements :}}
  The author is indebted to the authorities of IIT Roorkee for their unconditional support towards researches in basic sciences. The author would like to convey his sincere thanks to Prof. Carlos Nunez for several useful comments as well as pointing out a number of issues those are crucial for the analysis. The author would like to acknowledge The Royal Society, UK for financial assistance. The author would also like to acknowledge the Grant (No. SRG/2020/000088) received from The Science and Engineering Research Board (SERB), India. \\ 


\end{document}